\documentclass{article}

\usepackage{arxiv}
\usepackage[utf8]{inputenc} 
\usepackage[T1]{fontenc}    
\usepackage{url}            
\usepackage{booktabs}       
\usepackage{amsfonts}       
\usepackage{nicefrac}       
\usepackage{microtype}      
\usepackage{lipsum}		
\usepackage{graphicx}
\usepackage{titlesec}
\usepackage{achemso} 
\setkeys{acs}{usetitle = true}
\usepackage{multirow}
\usepackage{textcomp}

\usepackage{doi}
\usepackage{chngcntr}
\counterwithin{figure}{section}
\usepackage{ragged2e}
\usepackage{anyfontsize}
\usepackage{graphicx}
\usepackage{amsmath}
\usepackage{gensymb}
\usepackage{newtxtext} 
\usepackage{newtxmath} 
\usepackage{multicol}
\usepackage{hyperref}       
\usepackage{caption}        
\usepackage{xcolor}

\renewcommand{\footnotesize}{\fontsize{7.68pt}{9.22pt}\selectfont}
\renewcommand{\scriptsize}{\fontsize{6.72pt}{8.06pt}\selectfont}
\renewcommand{\tiny}{\fontsize{6.24pt}{7.49pt}\selectfont}
\renewcommand{\small}{\fontsize{8.64pt}{10.37pt}\selectfont}
\renewcommand{\normalsize}{\fontsize{10.24pt}{12.29pt}\selectfont}
\renewcommand{\large}{\fontsize{13.44pt}{16.19pt}\selectfont}
\renewcommand{\Large}{\fontsize{16.13pt}{19.43pt}\selectfont}
\renewcommand{\LARGE}{\fontsize{19.35pt}{23.22pt}\selectfont}
\renewcommand{\huge}{\fontsize{22.74pt}{27.29pt}\selectfont}
\renewcommand{\Huge}{\fontsize{27.29pt}{32.81pt}\selectfont}

\title{Intramolecular and water mediated \\ tautomerism of solvated glycine}

\author{
    \begin{center}
        \textbf{Pengchao Zhang}\textsuperscript{1,2}, \textbf{Axel Tosello Gardini}\textsuperscript{2,3}, \textbf{Xuefei Xu}\textsuperscript{1,*}, and \textbf{Michele Parrinello}\textsuperscript{2,*}.\\[0.5em]
        \textsuperscript{1}Center for Combustion Energy, Department of Energy and Power Engineering, and Key Laboratory for Thermal Science and Power Engineering of Ministry of Education, Tsinghua University, Beijing 100084, China\\[0.5em]
        \textsuperscript{2}Atomistic Simulations, Italian Institute of Technology, Genova 16152, Italy\\[0.5em]
        \textsuperscript{3}Department of Materials Science, Università di Milano-Bicocca, 20126 Milano, Italy\\[0.5em]
        \textsuperscript{*}Co-corresponding author e-mail: xuxuefei@tsinghua.edu.cn, michele.parrinello@iit.it
    \end{center}
}

\renewcommand{\headeright}{}
\renewcommand{\undertitle}{}
\renewcommand{\shorttitle}{\textit Zhang et. al., Intramolecular and water mediated tautomerism of solvated glycine}

\titlespacing{\section}{0pt}{\parskip}{-\parskip}
\titlespacing{\subsection}{0pt}{\parskip}{-\parskip}

\DeclareUnicodeCharacter{0308}{\"{o}}
\DeclareUnicodeCharacter{0301}{\'}

\begin{document}
\maketitle

\begin{abstract}
The understanding of prototropic tautomerism in water and the characterization of solvent effects on protomeric equilibrium pose significant challenges. Using molecular dynamics simulations based on state-of-the-art deep learning potential and enhanced sampling methods, we provide a comprehensive description of all configurational transformations in glycine solvated in water and determine accurate free energy profiles of these processes. We observe that the tautomerism between the neutral and zwitterionic forms of solvated glycine can occur by both intramolecular proton transfer in glycine and intermolecular proton transfer in the contact ion pair (anionic glycine and hydronium ion) or the separated ion pair (cationic glycine and hydroxide ion).
\end{abstract}

\begin{multicols}{2}
\renewcommand{\thefigure}{\arabic{figure}}
\justifying

\section*{1 Introduction}
\begin{figure*}
  \centering
  \includegraphics[width=\textwidth]{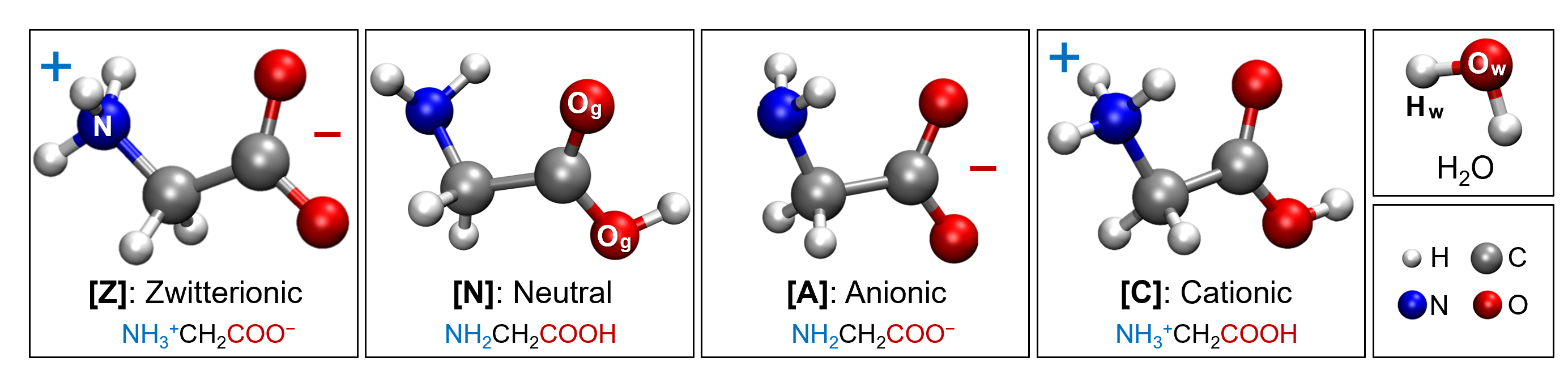}
  \caption{\textbf{The schematic diagrams of four glycine states and a water molecule.} [Z], [N], [A], and [C] denote zwitterionic, neutral, anionic, and cationic forms, respectively. Atomic notation is used to distinguish the atoms present in the molecules of glycine and water. $\mathrm{X_y}$ represents the atom X (H or O) that originates from molecule y (glycine or water).}
  \label{fig_4geo}
\end{figure*}

Prototropic tautomerism plays a crucial role in determining the thermodynamic and kinetic properties of amino acids and their derivatives, and consequently their reactivity and interaction with other biomolecules.\cite{elguero2000prototropic,singh2015role,pospisil2003tautomerism,raczynska2005tautomeric,douhal1995femtosecond,rodriquez2001proton,nemeria2009reaction} The simplest amino acid, glycine, can exist in the zwitterionic [Z], neutral [N], anionic [A] or cationic [C] form (Fig. \ref{fig_4geo}) depending on the pH of the solution\cite{sheinblatt1964nuclear,ottosson2011origins,hernandez2013protonation}, providing a simple yet relevant model for the study of tautomeric behavior. Here we will focus on the transformation of glycine in water between the [Z] and [N] forms. 

Experimental studies have been carried out to determine the protonation states of glycine under various conditions.\cite{locke1983effect,kumar2005vibrational,ottosson2011origins} It has been found that in the gas phase the [N] tautomer is the more stable. Under microsolvation conditions,\cite{alonso2006glycine,schwaab2022zwitter} it has been demonstrated that proton transfer can occur and the [Z] form is favoured. For fully solvated glycine,\cite{sheinblatt1964nuclear,locke1983effect,ottosson2011origins,hernandez2013protonation,sun2014understanding} the [Z] tautomer has also been shown to be more stable than the [N] tautomer. However, little is experimentally known on the dynamic processes of transformation from [N] to [Z], and these microscopic processes can be investigated by theoretical simulations at the atomic level.

The gas phase experiments have been accompanied by a number of static calculations.\cite{perez2016water,bachrach2008microsolvation,tripathi2021unveiling} However, the glycine tautomerism is difficult to achieve in the gas phase, but it is facilitated by the dynamics of the water environment and the fluctuation of the hydrogen bond (HB) network. In an effort to include water effects, various approaches have been taken, from describing water as polarizable continuum models (PCMs),\cite{wood2008nature,senn2003ab,aikens2006incremental,kayi2012theoretical,ottosson2011origins,hernandez2013protonation,valverde2018zwitterionization} to quantum mechanical/molecular mechanical (QM/MM) calculations,\cite{choi2012quantum,valverde2018zwitterionization} and to the use of reactive force fields.\cite{rahaman2011development} Most studies have come to the same conclusion as experiments, i.e., the [Z] tautomer is stable in water. However, only a few simulations have explored the tautomeric dynamics of glycine, and revealed the mechanism of short-range proton transfer, i.e., the transition from [N] to [Z] occurs intramolecularly or sometimes via a nearby solvent water 
molecule.\cite{rahaman2011development,choi2012quantum,tripathi2021unveiling} Here we aim at describing the static and dynamic behavior of glycine in water taking into account the full complexity of water dynamics and chemistry. The most appropriate theoretical framework for such a study is that of \textit{ab initio} molecular dynamics (AIMD).\cite{car1985unified,born1985quantentheorie,sun2010glycine,leung2005ab,tripathi2021unveiling}

Unfortunately, AIMD simulations are computationally expensive, and this has limited the size of the system studied and the time scale of the simulations. In pioneering work, Behler and Parrinello have shown that the cost of \textit{ab initio}-quality simulations can be greatly reduced without compromising accuracy.\cite{behler2007generalized} Their solution was to represent the potential energy surface as a suitably designed neural network whose inputs are a set of descriptors chosen so as to best describe the local atomic environment, while enforcing the symmetry of the problem and allowing scaling up to large systems. The neural network is then trained on a set of DFT (density functional theory) energies and forces performed on a carefully selected set of configurations. More recently, many variants of this approach have been proposed, such as Gaussian approximation potential,\cite{bartok2010gaussian} deep potential (DP),\cite{zhang2018endtoend,zhang2018deeppotential,wang2018deepmdkit} SchNet,\cite{schutt2018schnet} and other equivariant approaches.\cite{schutt2021equivariant,batzner20223,hoogeboom2022equivariant} These developments benefit greatly from easy access to efficient machine learning libraries and from efficient neural network training strategies. Here, we use the DP model to perform molecular dynamics (MD) simulations, which has proven its usefulness in many applications, including the studies of water and reactions in water.\cite{zhang2021phase,piaggi2022homogeneous,de2022acids,galib2021reactive,yang2022using,zhang2023double}

During the DP model training, it is important to present the network with an appropriate set of configurations, especially when we want to study reactive processes such as the protonation and deprotonation of glycine. Since these processes take place on a time scale that is much larger than the one accessible to standard MD simulations, the use of enhanced sampling methods is necessary to sample such rare events. Here we employ the recently developed on-the-fly probability enhanced sampling (OPES) method,\cite{invernizzi2020rethinking,invernizzi2022exploration} which is an evolution of metadynamics.\cite{laio2002escaping,barducci2008well} Like many other enhanced sampling methods, OPES uses collective variables (CVs) to accelerate the fluctuations that eventually lead to the reactive process. In this work we adapt the Voronoi CVs introduced by Grifoni et al.\cite{grifoni2019microscopic,grifoni2020tautomeric} so as to be able to simultaneously describe the [Z], [N], [A], [C] forms of glycine and the intermediate structures connecting these forms.  

Then we follow the deep Kohn-Sham (DeePKS)\cite{chen2020ground,chen2020deepks,chen2023deepkscpc,li2022deepksabacus} strategy to efficiently build a very accurate training dataset. In particular, we want to perform simulations in which the energies and forces are as accurate as those predicted by the hybrid meta-GGA functional M06-2X, which has been shown to give an accurate representation of the thermochemistry and kinetics of main group elements.\cite{zhao2008m06} However, the computational cost of performing single point calculations at the M06-2X level for the current system is too high. In simple terms, the way DeePKS gets around this difficulty is to train a neural network that expresses the difference $E_\delta$ between the target energy and the baseline energy. In our case the target energy is $E_{\mathrm{M06-2X}}$, and the baseline energy $E_{\mathrm{PBE}}$ is from the cheaper GGA functional PBE.\cite{perdew1996generalized} The $E_\delta$ is calculated as the sum of atomic energy differences, determined using a neural network that takes atomic coordinates, density matrices, and orbitals as inputs. The remarkable finding is that to train $E_\delta$ only a very small number of training data are needed. Once the PBE-based DeePKS model is trained, energy is calculated from $E_{\mathrm{PBE}}+E_\delta$, and forces can be driven from the energy. The energies and forces generated using such a DeePKS model will have an accuracy close to that of the full M06-2X calculation at a much lower cost, and will be used to train the DP model. 

We use the generated DP model to explore the tautomeric free energy landscape of glycine in water, and find that transitions between the [N] and [Z] forms of glycine can occur via multiple proton transfer (PT) processes.

\section*{2 Computational methods}
\subsection*{2.1 Potential model generation}

\subsubsection*{2.1.1 Building training datasets}
Key to the potential model training is the construction of appropriate training datasets. As discussed in the Introduction section, we need to train two potential models: One is the DP model to be used in molecular dynamics simulations for investigating the glycine tautomerism in water; the other is the PBE-based DeePKS model with the accuracy approaching the M06-2X functional to be used in the labeling of the DP training datasets. In either case we use pure water and solvated glycine configurations (see Table S1 in the SI). These configurations are generated in multiple independent MD simulations with enhanced sampling to generate as much uncorrelated data as possible and to include reactive processes. The details of the MD simulations are given below.

The numbers of configurations in the training sets of the two models are much different due to the difference between the descriptors. The similarities are that the two model descriptors all satisfy the physical symmetry and the locality. The differences are that the neural network descriptors in the DP model contain only the angular and radial atomic environment, while the descriptors of the DeePKS model additionally include density matrices projected on the atomic orbitals and satisfy the gauge invariance symmetry\cite{zhang2018deeppotential,chen2020deepks,li2022deepksabacus}. This means that only a few hundred configurations are needed to train the DeePKS model. In contrast, more than tens of thousands of configurations are needed for training the DP model to have an accurate modeling of the reactive process. Finally, a total of 300 and 55,498 configurations are collected in the DeePKS and DP model training datasets, respectively. 

\subsubsection*{2.1.2 DFT calculations}

To prepare DeePKS training datasets and all test datasets, the M06-2X\cite{zhao2008m06} energies and forces are calculated using the CP2K\cite{kuhne2020cp2k} package. In  the  calculations,  the Goedecker-Teter-Hutter pseudopotentials\cite{goedecker1996separable,hartwigsen1998relativistic} are used together with a quadruple-zeta valence basis set with polarization\cite{vandevondele2005quickstep} (QZV3P). The multi-grid level utilizes a plane-wave cutoff of 1000 Ry for the total density and 70 Ry for the Kohn-Sham orbitals. To accelerate the convergence of self-consistent field (SCF) iterations, the auxiliary density matrix method\cite{guidon2010auxiliary} is utilized.

\subsubsection*{2.1.3 DeePKS model training}

The DeePKS model is then generated using an iterative approach, in which we alternately train the network of the correction term from the PBE baseline to the M06-2X target using the DeePKS-kit \cite{chen2020ground,chen2020deepks,chen2023deepkscpc} package, and solve the resulting DeePKS model in ABACUS\cite{chen2010systematically,li2016large,li2022deepksabacus}.  In the ABACUS calculations, the optimized norm-conserving Vanderbilt\cite{schlipf2015optimization} pseudopotentials are used together with numerical atomic orbital\cite{lin2021strategy} basis. The kinetic energy cutoff is set at $100$ Ry, and the SCF convergence threshold for the density error is $1\times 10^{-7}$ Ry.

The above iterations are no longer needed once the DeePKS model has been trained. One performs single point calculations based on the PBE functional corrected with the $E_\delta$ term to obtain total energies and forces. Relative to an independent set of  testing data obtained by performing ordinary M06-2X/QZV3P calculations,  the PBE-based DeePKS model shows root mean square errors (RMSEs) of 0.53 and 0.61 meV/atom for the energies and of 43 and 52 meV/{\AA} for the forces, where we distinguish between errors relative to pure water and solvated glycine systems.  In terms of efficiency, the PBE-based DeePKS model saves about an order of magnitude in time compared to using the standard M06-2X functional (Table S3 in the SI). 

\subsubsection*{2.1.4 DP model training}

The DP model is then trained in the DeePMD-kit\cite{wang2018deepmdkit,zeng2023deepmd} package using the training dataset of M06-2X quality generated as described above.  As discussed earlier, at this stage we needed 55,498 configurations for an accurate result. Then, the trained DP model is examined by testing the prediction on an independently generated dataset containing 6,020 configurations for which standard M06-2X energy and force calculations have been performed.  The DP model exhibits energy RMSEs of 0.79 and 1.72 meV/atom and atomic force RMSEs of 58 and 63 meV/{\AA} for water and solvated glycine systems, respectively, suggesting that the DP model has achieved a level of precision comparable to M06-2X. Compared to an M06-2X functional based simulation, the DP model is five orders of magnitude faster (Table S3). 

\subsection*{2.2 Molecular dynamics simulations}

\subsubsection*{2.2.1 MD simulation for the training dataset generation} 

The configurations required to train the DeePKS and DP models are sampled from MD simulations using CP2K\cite{kuhne2020cp2k} as the driver. To speed up the sampling, we use the semi-empirical GFN1-xTB\cite{grimme2017robust} to derive potential energy surface. This GFN1-xTB method has a lower computational cost, but is still accurate enough to describe the structures of different glycine forms in water. The simulations are performed in constant volume constant temperature (NVT) ensemble with a time step of 1.0 fs. The temperature of 300 K is enforced using the velocity rescaling thermostat\cite{bussi2007canonical} with a damping time of 0.2 ps. 

\subsubsection*{2.2.2 DPMD simulation for studying glycine tautomerism}

To study the glycine tautomerism in water, deep potential molecular dynamics (DPMD) simulations are performed using the LAMMPS \cite{plimpton1995fast} software. Before the formal simulation, we have investigated the size effect by placing a glycine molecule in varying amounts of water. Based on the convergence test result (Fig. S9), the system C (Table S1) with one glycine molecule in 128 H$_2$O molecules is used in the final DPMD simulation to strike a balance between efficiency and precision. In order to sufficiently sample all possible tautomeric processes, simulations are run for 30 ns in the NVT ensemble with an integration time step of 1.0 fs. The temperature is controlled at 300 K using again the velocity rescaling thermostat with a damping time of 0.04 ps. 

\subsection*{2.3 Enhanced sampling settings}

\subsubsection*{2.3.1 Bias potential}

The OPES method using collective variables (CVs) is employed in both MD simulations for constructing the datasets and the DPMD simulation of glycine tautomerism. The CVs $\mathbf{{s}({R})}$, which are functions of atomic coordinates ($\mathbf{{R}}$), skilfully capture the slow modes associated with rare events. During MD simulations, the potential energy ($U(\mathbf{R})$) of the system is modified by adding the external bias potential ($V(\mathbf{s})$) through the utilization of the PLUMED\cite{tribello2014plumed} plugin. Specifically, $V(\mathbf{s})$ at the $n$th step in OPES\cite{invernizzi2020rethinking,invernizzi2022exploration} is defined by:
\begin{equation}
    V_n(\mathbf{s}) = (1-{\frac{1}{\gamma}})\frac{1}{\beta} \log (\frac{{P}_{n}{(\mathbf{s})}}{Z_n}+\epsilon)
\label{eq:V(s)}
\end{equation}
where $\beta = 1/{k_B T}$ is the inverse Boltzmann factor. Probability $P(\mathbf{s})$ is the unbiased marginal distribution, and parameter $Z$ is the normalization factor.  Bias factor $\gamma = \beta \Delta E_{\mathrm{bias}}$ ensures the reshaping of the original probability. The regularization term $\epsilon= e^{-\gamma/(1-1/\gamma)}$ not only guarantees a positive argument for the logarithm, but also imposes a bias constraint, thereby limiting sampling within the specified region of interest. In particular, the bias update is carried out every 500 steps with the adaptive kernel width, and the value of $\Delta E_{\mathrm{bias}}$ is set at 35 kJ/mol. 

Notably, the "explore" variant of OPES (OPES-explore) is used in the configuration collection of datasets due to its ability to accelerate the exploration of phase space, while the OPES is performed to generate well-converged free energies during the final DPMD simulation. The reason is that the ways of estimating the probability distribution are different in the OPES and OPES-explore methods, although the idea of defining the bias potential is similar. Specifically, unbiased probability $P(\mathbf{s})$ is estimated on-the-fly using weighted kernel density estimation (KDE) in OPES, while the well-tempered probability $p^{\mathrm{WT}}(\mathbf{s}) (\propto [P(\mathbf{s})]^{\frac{1}{\gamma}}$) is estimated based on averaged KDE in OPES-explore.\cite{invernizzi2020rethinking,invernizzi2022exploration}

\subsubsection*{2.3.2 Collective variables}

Proton can diffuse through water via the Grotthuss mechanism,\cite{von1805memoire,marx2006proton,zhang2023double} in which it is not a well-specified proton that moves, but rather a charge defect that migrates through the water hydrogen network. Thus a CV that is used to describe proton diffusion has to be able to identify charge defects without making reference to a specific set of atomic coordinates. In the present case, several charge defects are possible: the --NH$_3$$^+$ and --COO$^-$ groups in glycine, and the two water self-ions (hydronium H$_3$O$^+$ and hydroxide OH$^-$). To identify automatically these charge defects we follow the strategy of References\cite{grifoni2019microscopic,grifoni2020tautomeric} and tessellate the space with Voronoi polyhedra centered on the O and N atoms. We then sum the charge contained in each polyhedron. If the charge in a polyhedron is different from zero, we attribute the charge defect to the atom N or O that is at the polyhedron center, and we distinguish the O defects according to whether they are centered on water or glycine oxygen ($\mathrm{O_w}$ and $\mathrm{O_g}$ in Fig. \ref{fig_4geo}). 

To count the number of H atoms $n_i$ centered on O or N  atom $i$, we use the formula 
\begin{equation}
    n_i = \sum_{j=1}^{\mathrm{Num_H}} \frac{e^{-\lambda |\mathbf{R}_i - \mathbf{R}_j|}}{\sum_{m=1}^{\mathrm{Num_{O\&N}}} e^{-\lambda |\mathbf{R}_m - \mathbf{R}_j|}}
\label{eq:n_i}
\end{equation}
where the first sum is over all H atoms with index $j$, and $m$ is the index of Voronoi centers. The parameter $\lambda$ regulates the smoothness of the function. 

The charge defect number $\delta_i$ of the Voronoi center $i$ is calculated by subtracting the reference number $n_i^0$ (the original proton number connecting to the atom at the polyhedron center) from H number $n_i$ 
\begin{equation}
    \delta_i = n_i - n_i^0
\label{eq:delta_i}
\end{equation}
where we take $n_i^0=2$ for the water oxygen ($\mathrm{O_w}$) and the glycine nitrogen (N), while for the carboxylic oxygens ($\mathrm{O_g}$) we set $n_i^0=\frac{1}{2}$ on account of the symmetry between two oxygen atoms. 

As the system is neutral overall, the [C] and [A] forms of glycine are compensated by OH$^-$ and H$_3$O$^+$, respectively. Then, to distinguish between the [C]$-$OH$^-$ pair, [Z]$\&$[N] states and [A]$-$H$_3$O$^+$ pair, we define a CV $\mathbf{s}_p$ that can identify these protonation states, by combining the charge defects of both glycine and water. 
\begin{equation}
    \mathbf{s}_p = \sum_{i=1}^{\mathrm{Num_{O_w}}} \delta_{i} + 2 (\sum_{j=1}^{\mathrm{Num_{N}}} \delta_{j} + \sum_{k=1}^{\mathrm{Num_{O_g}}} \delta_{k})
\label{eq:s_p}
\end{equation}
where $\mathbf{s}_p \approx 1$ for [C]$-$OH$^-$,  $\mathbf{s}_p \approx -1$ for [A]$-$H$_3$O$^+$,  and $\mathbf{s}_p \approx 0$ in the other two cases. 
Since $\mathbf{s}_p$ cannot distinguish the [Z] and [N] forms,  we introduce another CV $\mathbf{s}_d$ to estimate the charge-charge distance that is capable of separating the two cases. The $\mathbf{s}_d$ is measured in terms of the distances between $\mathrm{O_w}$ and N and between $\mathrm{O_w}$ and $\mathrm{O_g}$, as well as the average distance between N and $\mathrm{O_g}$, as follows:
\begin{equation}
\begin{split}
\mathbf{s}_d = & -\sum_{i=1}^{\mathrm{Num_{O_w}}} \sum_{j=1}^{\mathrm{Num_{N}}} r_{i,j} \delta_{i} \delta_{j} - 
\sum_{i=1}^{\mathrm{Num_{O_w}}} \sum_{k=1}^{\mathrm{Num_{O_g}}} r_{i,k} \delta_{i} \delta_{k} \\
& - \sum_{j=1}^{\mathrm{Num_{N}}} \sum_{k=1}^{\mathrm{Num_{O_g}}} r_{j,k} \delta_{j} \delta_{k}
\end{split}
\label{eq:s_d}
\end{equation}
where $r$ is the modulus distance between the two Voronoi centers. The CV $\mathbf{s}_d$ will approximately be 0 and 3 {\AA} in the [N] and [Z] forms, respectively, and will become larger than \textasciitilde 3 {\AA} in the other two cases.

During the OPES simulations, both CVs $\mathbf{s}_p$ and $\mathbf{s}_d$ use the value of $\lambda=5$ so as to have a smoother definition of the Voronoi polyhedra.

Since H$_3$O$^+$ and OH$^-$ can be present in water,  we have added configurations related to the autoionization process of water into our training datasets.  In the potential training process we use the CV 
\begin{equation}
    \mathbf{s}_a = \sum_{i=1}^{\mathrm{Num_{O_w}}} \delta_{i}^2 \
\label{eq:s_a}
\end{equation}
to promote the self ionization processes, and $\mathbf{s}_a$ varies from 0 to 2. 

The CV $\mathbf{s}_a$ is supplemented by a CV  $\mathbf{s}_t$  that represents the distance between H$_3$O$^+$ and OH$^-$: 
\begin{equation}
    \mathbf{s}_t = -\sum_{i=1}^{\mathrm{Num_{O_w}}} \sum_{j>i}^{\mathrm{Num_{O_w}}} r_{i, j} \delta_{i} \delta_{j} \
\label{eq:s_t}
\end{equation}
While $\mathbf{s}_t$ can distinguish between the pure water state ($\mathbf{s}_t \approx 0$) and autoionization state ($\mathbf{s}_t>0$), it is hardly to identify the initial proton transfer which corresponds to a  $\mathbf{s}_t$ value nearly concentrated around zero.  Therefore, we use a  piecewise logarithmic function:
\begin{equation}
    \mathbf{s}'_{t} = \begin{cases}
\log(\mathbf{s}_t + \epsilon), & \ 0 \leq \mathbf{s}_t < 1 \\
\mathbf{s}_t - 1 + \log(1+\epsilon), & \ \mathbf{s}_t \geq 1
\end{cases}
\label{eq:s_logt}
\end{equation}
where $\epsilon = 0.03$ is a regularization parameter.

\subsubsection*{2.3.3 Free energy calculation}
After the DPMD simulation, the free energy surface (FES) along a given CV can be calculated as follows:
\begin{equation}
F(\mathbf{s}) = -\frac{1}{\beta} \log P(\mathbf{s})
\label{eq:F(s)}
\end{equation}
In the regime where the bias is quasi-static, $P(\mathbf{s})$ can be reweighted\cite{invernizzi2020rethinking} as an average over the biased ensemble.
\begin{equation}
P(\mathbf{s}) = \frac{\langle \delta[\mathbf{s} - \mathbf{s}(\mathbf{R})]e^{\beta V(\mathbf{s})} \rangle_V}{\langle e^{\beta V(\mathbf{s})} \rangle_V}
\label{eq:P(s)}
\end{equation}
See the supporting information for the full simulation workflow and more computational details.

\section*{3 Results and discussion}
\subsection*{3.1 Free energy surfaces}

\begin{figure*}
  \centering
  \includegraphics[width=\textwidth]{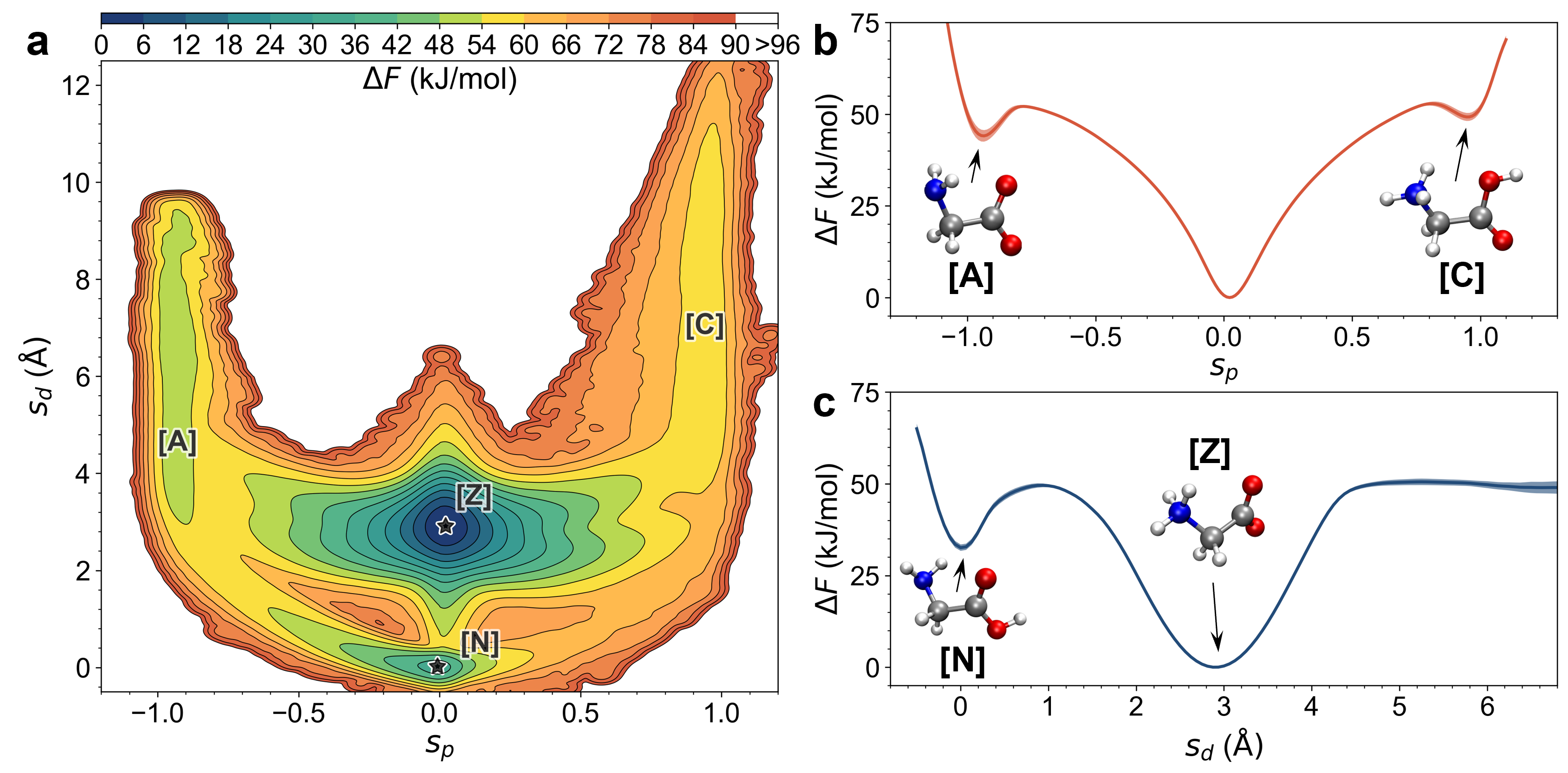}
  \caption{\textbf{Free energy profiles.} \textbf{a}. Two-dimensional FES as a function of the glycine protonation coordinate $\mathbf{s}_p$ and the charge-charge distance $\mathbf{s}_d$. \textbf{b} and \textbf{c}. One-dimentional projections of the FES along the CVs $\mathbf{s}_p$ and $\mathbf{s}_d$. The standard deviations of one-dimentional FESs are represented by transparent colors.} 
  \label{fig_1d2dfes_geo}
\end{figure*}

The converged FES is plotted in Fig. \ref{fig_1d2dfes_geo}a as a function of the two CVs $\mathbf{s}_p$ and $\mathbf{s}_d$, which reflect the glycine protonation state and the charge-charge distance, respectively. The minima corresponding to state [N] and state [Z] are easily identified. Less evident is the presence of two other metastable states, [A] and [C]. Their existence can be made more clear if we project the free energy along the CV $\mathbf{s}_p$ as shown in Fig. \ref{fig_1d2dfes_geo}b. In this one-dimensional representation, the [N] and [C] forms cannot be resolved and are part of one single central minimum, but clearly two local minima that correspond to [A] and [C] can be detected, and these states lie higher in energy relative to the minimum by 44.1 kJ/mol ([A]) and 49.2 kJ/mol ([C]), respectively. To facilitate reading the result in a one dimensional projection, we also plot the FES along the CV $\mathbf{s}_d$ (Fig. \ref{fig_1d2dfes_geo}c). In this projection, [A] and [C] cannot be identified, while the [Z] and [N] states are now clearly visible. The energy difference between the lowest free energy [Z] state and the [N] state is 32.6 kJ/mol, in good agreement with experiments, which have reported values ranging from 30.4 kJ/mol to 32.1 kJ/mol.\cite{wada1982ratio,haberfield1980energy,slifkin1984thermodynamic}  

It is interesting to analyze in some detail the nature of the [A] and [C] states. Because of the requirement that the system must be neutral, the charged glycine protomers in these two states are accompanied by a counterion, which is H$_3$O$^+$ in [A] and OH$^-$ in [C]. The behavior of these two ion pairs is different; in [A] the hydronium remains close to the glycine, while in [C] the ion pair can separate more easily. This different behavior is reflected in the different $\mathbf{s}_d$ distributions. The $\mathbf{s}_d$ distribution in [C]$-$OH$^-$ pair is much broader than that of [A]$-$H$_3$O$^+$ pair, and as a consequence the FES exhibits a slight left-right asymmetry of Fig. \ref{fig_1d2dfes_geo}a. 

\subsection*{3.2 Prototropic tautomerism pathways}

\begin{figure*}
  \centering
  \includegraphics[width=\textwidth]{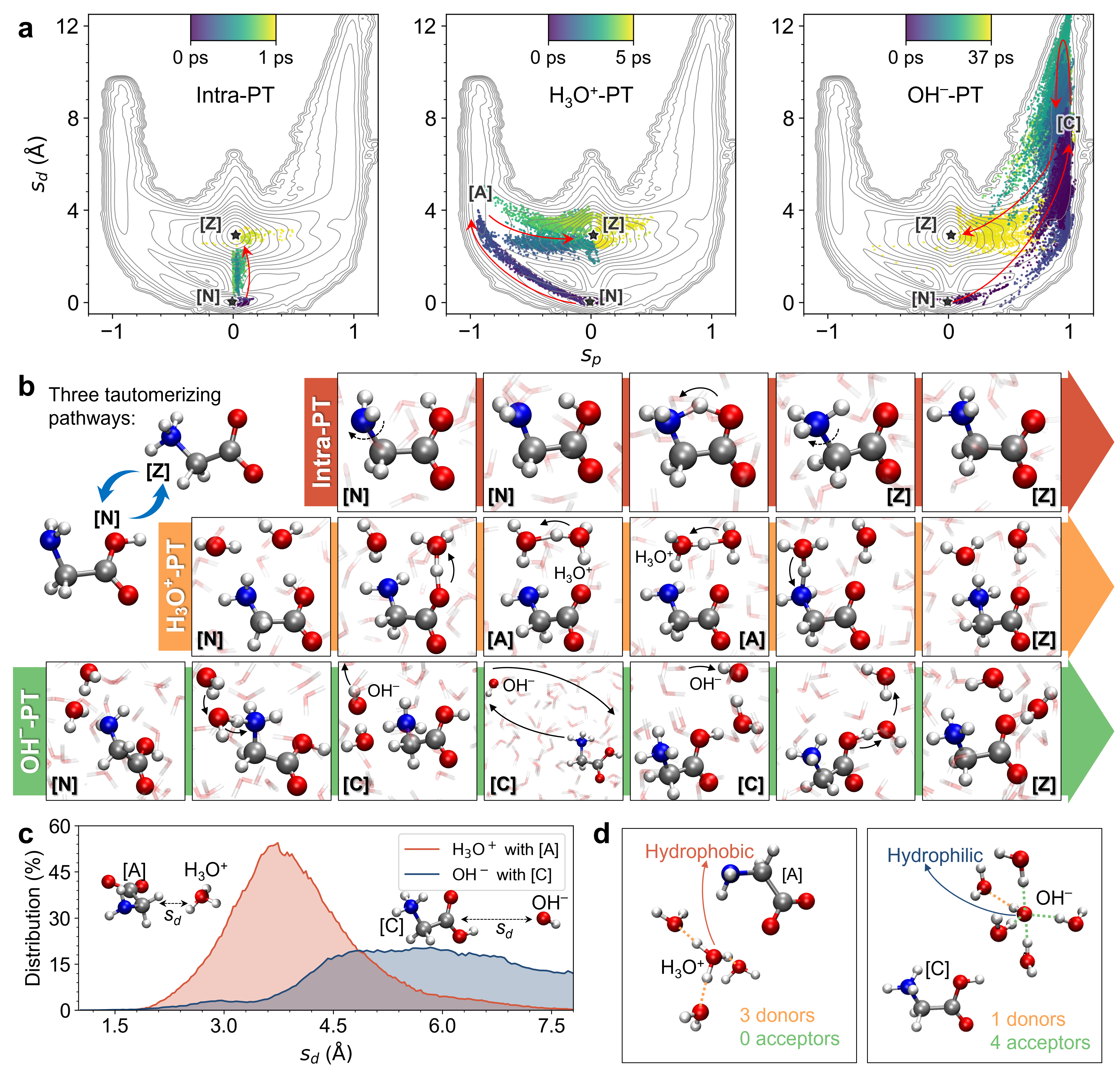}
  \caption{\textbf{Three tautomeric pathways.} \textbf{a}. Continuous sampling processes of tautomeric pathways with the CVs $\mathbf{s}_p$ and $\mathbf{s}_d$, where the color bar represents the relative simulation time. \textbf{b}. Trajectories of [N]$-$[Z] tautomerism along the Intra-PT, H$_3$O$^+$-PT, and OH$^-$-PT pathways. The molecular configurations involved in the reaction pathways are highlighted in the sphere models, and other surrounding water molecules are shown in the transparent stick models. \textbf{c}. The distribution of the CV $\mathbf{s}_d$, i.e., intermolecular charge-charge distance (no intramolecular distance in the [A] and [C] cases). \textbf{d}. The hydrogen-bond schematic diagrams of H$_3$O$^+$ and OH$^-$ ions near [A] and [C].}
  \label{fig_3path}
\end{figure*}


We now analyze the pathways leading from [N] to [Z]. We have identified three possible proton transfer pathways as shown in Fig. \ref{fig_3path}a,b. The most obvious is a direct intramolecular proton transfer (Intra-PT), which is consistent with the gas phase studies \cite{jensen1995number,tang2018development,tunon2000tautomerization}. The --NH$_2$ group first rotates so that its lone pair helps to accept the --COOH proton to form the ammonium group --NH$_3$$^+$, which later rotates to assume a more stable [Z] conformation.

In addition to the direct route, there are two other pathways in which the proton is transferred via a Grotthuss-like mechanism\cite{von1805memoire,marx2006proton,zhang2023double} with the aid of the water molecules in the solvent. These two mechanisms have either [A] or [C] as an intermediate; in the first case, a contact ion pair $[\mathrm{A}]-\mathrm{H}_3\mathrm{O}^+$ is formed, whereas in the second one passes through the separated ion pair $[\mathrm{C}]-\mathrm{OH}^-$.

In the anionic pathway, the carboxylic proton is transferred to a nearby solvent water, which transfers a positive charge to another water molecule via a Zundel intermediate to end up on the glycine, forming the ammonium group of the [Z] state. This $\mathrm{H_3O^+}$ mediated proton transfer ($\mathrm{H_3O^+}$-PT) process is also found at the microhydration limit of glycine-water clusters.\cite{jensen1995number,tripathi2021unveiling,tang2018development}

Similar but different is the new pathway that leads from [N] to [Z] via an intermediate [C]$-$OH$^-$ pair mediated proton transfer (OH$^-$-PT). The first step is the transfer of a proton from a water molecule to the amine group of glycine to form an ammonium cation and an OH$^-$ ion. The OH$^-$ ion is then solvated  in  water  and diffuses a relatively long distance via a Grotthuss mechanism, eventually ending up back at the glycine to abstract the carboxylic proton to form the [Z] protomer and a water molecule. 

The reason for the distinct charge-charge distance behavior comes from the different amphipathy of $\mathrm{H_3O^+}$ and $\mathrm{OH^-}$ ions as depicted in Fig. \ref{fig_3path}c,d, where the distance between the contact ion pair $[\mathrm{A}]-\mathrm{H}_3\mathrm{O}^+$ is in the short range (< \textasciitilde4.5 
 {\AA}), while the distance between the separated ion pair $[\mathrm{C}]-\mathrm{OH}^-$ is in the relatively long range (> \textasciitilde4.5 
 {\AA}). Since glycine disrupts the HB network of water and provides a hydrophobic environment, $\mathrm{H_3O^+}$ tends to stay in proximity to glycine due to its hydrophobic O atom. Compared to $\mathrm{H_3O^+}$, the O atom in $\mathrm{OH^-}$ is hydrophilic and can easily form HB with surrounding water molecules as HB acceptors,\cite{chen2018hydroxide,zhang2023double} resulting in a more extensive HB network around $\mathrm{OH^-}$. This indicates that $\mathrm{OH^-}$ can diffuse further away from glycine towards the outer water solvation shell compared to $\mathrm{H_3O^+}$, and thus the outer solvation shell of glycine is involved in the prototropic tautomerism and cannot be ignored. 

Note that experimental investigations\cite{slifkin1984thermodynamic} on the glycine protonation reactions were mostly conducted in the water with glycine concentrations falling within the interval of 0.03-0.25 M at room temperature. The present simulation results are obtained based on a neutral aqueous simulation system with a little bit higher glycine concentration of 0.43 M. Nevertheless, as documented in the literature, the solubility of glycine in water spans a range of 0.36-0.56 M,\cite{bowden2018solubility} and therefore the present simulation ensures a solvated glycine molecule. In addition, it is noteworthy that the discovery of reaction pathways involving ion pairs in the simulation, accompanied by the presence of $\mathrm{H_3O^+}$ or $\mathrm{OH^-}$ in the aqueous environment, indicates an instantaneous change of the pH, leading to values of 0.4 or 13.6, respectively. Under the present simulated condition, we have clearly delineated three distinct [N]$-$[Z] tautomerization pathways for the first time, elucidated the roles and characters of [C]$-$OH$^-$ and [A]$-$H$_3$O$^+$ ion pairs in these processes, and indicated a way by which the pH of the solution can influence the protomeric equilibrium. This potentially have broader applicability in chemistry, biology, engineering and other proton transfer processes. We look forward to the upcoming advanced experiments that will serve to validate our findings. It is also important to emphasize that the quantitative results of these experiments will depend on the glycine concentration and, more interestingly, the pH of the system.

\section*{4 Conclusion}
In summary, the present study provides novel insights into the prototropic tautomerism of glycine in water and covers all possible configurational transformations using an accurate description of the interaction potential and a thorough sampling of the potential energy surface. We discover three pathways for tautomerization between the neutral and zwitterionic forms of solvated glycine; one is via intramolecular proton transfer in glycine, the second one involves short-range intermolecular proton transfer in the contact ion pair between anionic glycine and hydronium ion, and the third one has the aid of long-range intermolecular proton transfer in the separated ion pair between cationic glycine and hydroxide ion. In the two intermolecular proton transfer pathways, the observed remarkably distinct charge-charge distance of the intermediate ion pairs is attributed to the different amphipathy of water self-ions.

The combination of our computational technologies, including DeePKS, DeePMD, OPES, and Voronoi CVs, not only deepens our understanding of glycine tautomerism in water, but also provides a comprehensive framework and methodology for facilitating further research into the intricate dynamics of proton transfer.

\section*{Acknowledgments}
This work received partial support from the National Natural Science Foundation of China under Grant No. [21973053]. The authors would like to express their gratitude to Umberto Raucci, Enrico Trizio, Sudip Das, Linfeng Zhang, Qi Ou, Andrea Rizzi and Francesco Mambretti for their valuable discussions. Computational resources were provided by the High Performance Computing (HPC) platform at Tsinghua University and the HPC Franklin at Fondazione Istituto Italiano di Tecnologia.

\section*{Author contributions}
All authors contributed to the conception and design of the study. P.Z. and A.T.G. conducted the testing of the enhanced sampling method. P.Z. performed the simulation workflow and data analysis, with the assistance of all authors. X.X. and M.P. provided supervision throughout stages of the project. The initial draft of the manuscript was prepared by P.Z., and all authors participated in editing and reviewing the final version.

\renewcommand{\bibfont}{\footnotesize\linespread{0.9}\selectfont}
\bibliography{references}
\end{multicols}
\newpage

\section*{\centering Table of contents (TOC)}

\begin{figure}
  \centering
  \includegraphics[width=0.6\textwidth]{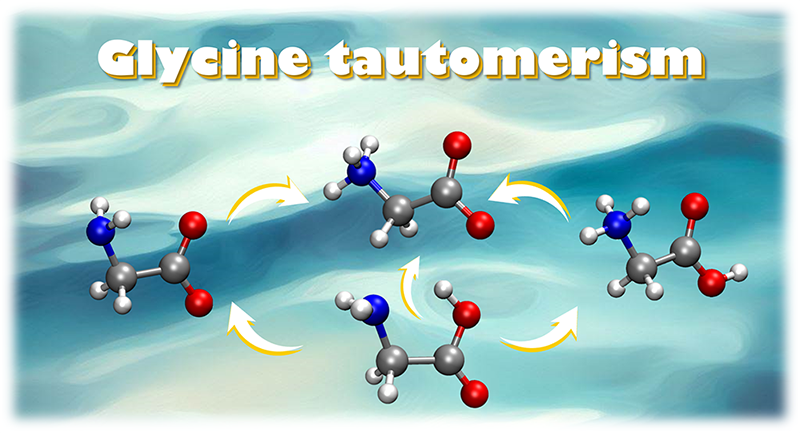}
  \label{fig_toc}
\end{figure}

\end{document}


\maketitle

\renewcommand{\thepage}{SI-\arabic{page}}
\renewcommand{\thefigure}{S\arabic{figure}}
\renewcommand{\thetable}{S\arabic{table}}
\renewcommand{\theequation}{S\arabic{equation}}
\justifying

\section*{DeePKS model training}

The DeePKS model employs descriptors derived from the projected density matrices, necessitating a predefined set of projectors with a maximum angular momentum of 2. The number of Bessel functions is determined by the radial and wavefunction cutoffs, specifically 5 Bohr and 100 Ry, respectively. The number of neurons is set to $[32, 32, 32]$ in hidden layers and the non-linear activation function between hidden layers is "gelu". The train step is set to $10$ thousand with a learning rate of $10^{-4}$ to $10^{-7}$. A weighted mean square loss function $\mathcal{L}^\mathrm{DeePKS}$ is applied to energy and atomic forces. 
\begin{equation}
    \mathcal{L}^\mathrm{DeePKS}=\left|E^{\mathrm{M06-2X}}-E^{\mathrm{DeePKS}}\left(\varphi_i | w\right)\right|^{2}+p\left|\mathbf{F}^{\mathrm{M06-2X}}-\mathbf{F}^{\mathrm{DeePKS}}\left(\varphi_i | w\right)\right|^{2}
\label{si_eq:loss_ks}
\end{equation}
Energy and force are determined by the Hamiltonian. The total Hamiltonian of the DeePKS model ($\hat{H}^{\mathrm{DeePKS}}$) is expressed as follows:
\begin{equation}
\hat{H}^{\mathrm{DeePKS}} = \hat{H}^{\mathrm{PBE}} + \hat{H}^{\delta}
\label{eq:H_DeePKS}
\end{equation}
where $\hat{H}^{\mathrm{DeePKS}}$ incorporates a correction term ($\hat{H}^{\delta}$) that is added onto the baseline functional ($\hat{H}^{\mathrm{PBE}}$). $\varphi_i$ is the eigenstate of Hamiltonian $\hat{H}^{\mathrm{DeePKS}}$ depending on the correction term $\hat{H}^{\delta}$. $\hat{H}^{\delta}$ is determined by projected density matrices, localized orbitals, and the DeePKS descriptor. $\hat{H}^{\mathrm{DeePKS}}$ after adding $\hat{H}^{\delta}$ corresponds to different ground states, so the model training and SCF solution are performed in turn, iterating until convergence.\cite{chen2020deepks,li2022deepksabacus} $\hat{H}^{\mathrm{DeePKS}}$ is approximately equal to the M06-2X Hamiltonian ($\hat{H}^{\mathrm{M06-2X}}$) when the wave function is converged, then the corresponding energy and force can be solved. In addition, $w$ is the neural network parameter. $p$ is the pre-factor to balance the errors.

\section*{DP model training}
The descriptor of Deep Potential Smooth Edition (DeepPot-SE) is used\cite{zhang2018endtoend,zhang2018deeppotential}, where the embedding net size is set to $[25, 50, 100]$ per layer and the non-linear activation function between hidden layers is "tanh". The submatrix size in the embedding net is set to 16. A neighbor atom cutoff radius of $6.0$ {\AA} with smoothing beginning at $0.5$ {\AA}, and the maximum neighbor number of $[\mathrm{H, O, N, C}]$ is $[85, 45, 1, 2]$. The fitting net is connected after the embedding net, with a size of $[240, 240, 240]$ per layer and the non-linear activation function between hidden layers is "tanh". The ResNet \cite{he2016deep} architecture is built between the fitting net. The train step is set to $10$ million with a learning rate of $10^{-3}$ to $10^{-8}$. A weighted mean square loss function $\mathcal{L}^\mathrm{DP}$ is applied to energy and atomic forces. 
\begin{equation}
    \mathcal{L}^\mathrm{DP}=\frac{p_{e}}{N}\left|E^{\mathrm{DeePKS}}-E^{\mathrm{DP}}\left(w\right)\right|^{2}+\frac{p_{f}}{3 N} \sum_{i \alpha}\left|F_{i \alpha}^{\mathrm{DeePKS}}-F_{i \alpha}^{\mathrm{DP}}\left(w\right)\right|^{2}
\label{si_eq:loss_dp}
\end{equation}
where $N$ is the number of atoms. $p_e$ and $p_f$ are the pre-factor of energy $E$ and force $F$, respectively. $w$ is the neural network parameter. $F_{i \alpha}$ means the force of the atom $i$ along the $\alpha$th direction. 

\section*{Derivatives of collective variables}
When adapting Voronoi CVs ($\mathbf{s(R)}$) to distinguish glycine states, it is important to note that these CVs can be derived with respect to the atomic coordinates ($\mathbf{R}$), as indicated in the following equation \ref{si_eq:d_V_s}:
\begin{equation}
    \frac{\partial V\mathbf{(s)}}{\partial \mathbf{R}}=\frac{\partial V\mathbf{(s)}}{\partial \mathbf{s(R)}} \frac{\partial \mathbf{s(R)}}{\partial \mathbf{R}}
\label{si_eq:d_V_s}
\end{equation}
where the implementation of the first term $\frac{\partial V\mathbf{(s)}}{\partial \mathbf{s(R)}}$ has been completed within the PLUMED \cite{tribello2014plumed} plugin, we will proceed to describe the second term $\frac{\partial \mathbf{s(R)}}{\partial \mathbf{R}}$ in detail and make corresponding changes to the code. For clarity and ease of understanding, we will focus on the relatively complicated part of the Voronoi CVs, as represented by the following equation \ref{si_eq:cv_s}. If there is a need to obtain derivatives for other CVs ($\mathbf{s}_p$, $\mathbf{s}_d$, $\mathbf{s}_a$, and $\mathbf{s}_{t}$), such calculations can be easily achieved by applying the derivative chain rule.
\begin{equation}
    \mathbf{s(R)}=\sum_{i \in \text{group}}{\delta_i}
\label{si_eq:cv_s}
\end{equation}
Next, we will illustrate the process using the example of distinguishing autoionization in pure water without any prior information. In this scenario, all O atoms act as Voronoi centers and form a defined group. To identify the protonation states of all O atoms, all nearby H atoms around the target O$_i$ atom (labeled as O$_i\rightarrow$H$_\mathrm{all}$) and all nearby O atoms around the target H$_j$ atom (labeled as O$_i\rightarrow$H$_j\rightarrow$O$_\mathrm{all}$) are searched.

In O$_i\rightarrow$H$_\mathrm{all}$ stage, if H$_j$ (selected in H$_\mathrm{all}$) is a neighbor of a given target O$_i$, the derivative for H$_j$ coordinate is governed by the following equation \ref{si_eq:d_s_H_1}. For simplicity, some small terms are ignored.
\begin{equation}
    \frac{\partial \mathbf{s(R)}}{\partial \mathbf{R}_{\mathrm{H}_j}} \approx \sum_{i \in \mathrm{O}} \sum_{j \in \mathrm{H}}-\lambda \frac{e^{-\lambda\left|\mathbf{R}_{\mathrm{O}_i}-\mathbf{R}_{\mathrm{H}_j}\right|}}{\sum_{m \in \mathrm{O}} e^{-\lambda\left|\mathbf{R}_{\mathrm{O}_m}-\mathbf{R}_{\mathrm{H}_j}\right|}}\left[1-\frac{e^{-\lambda\left|\mathbf{R}_{\mathrm{O}_i}-\mathbf{R}_{\mathrm{H}_j}\right|}}{\sum_{m \in \mathrm{O}} e^{-\lambda\left|\mathbf{R}_{\mathrm{O}_m}-\mathbf{R}_{\mathrm{H}_j}\right|}}\right] \frac{\mathbf{R}_{\mathrm{H}_j}-\mathbf{R}_{\mathrm{O}_i}}{\left|\mathbf{R}_{\mathrm{O}_i}-\mathbf{R}_{\mathrm{H}_j}\right|}
\label{si_eq:d_s_H_1}
\end{equation}
if H$_j$ (selected in H$_\mathrm{all}$) is a neighbor of a non-target Voronoi center O$_n$ ($n \neq i$), the derivative for H$_j$ coordinate is given by the following equation \ref{si_eq:d_s_H_0}. For simplicity, some small terms are ignored.
\begin{equation}
    \frac{\partial \mathbf{s(R)}}{\partial \mathbf{R}_{\mathrm{H}_j}} \approx \sum_{i \in \mathrm{O}} \sum_{j \in \mathrm{H}}-\lambda \frac{e^{-\lambda\left|\mathbf{R}_{\mathrm{O}_n}-\mathbf{R}_{\mathrm{H}_j}\right|}}{\sum_{m \in \mathrm{O}} e^{-\lambda\left|\mathbf{R}_{\mathrm{O}_m}-\mathbf{R}_{\mathrm{H}_j}\right|}}\left[0-\frac{e^{-\lambda\left|\mathbf{R}_{\mathrm{O}_i}-\mathbf{R}_{\mathrm{H}_j}\right|}}{\sum_{m \in \mathrm{O}} e^{-\lambda\left|\mathbf{R}_{\mathrm{O}_m}-\mathbf{R}_{\mathrm{H}_j}\right|}}\right] \frac{\mathbf{R}_{\mathrm{H}_j}-\mathbf{R}_{\mathrm{O}_n}}{\left|\mathbf{R}_{\mathrm{O}_n}-\mathbf{R}_{\mathrm{H}_j}\right|}
\label{si_eq:d_s_H_0}
\end{equation}

In O$_i\rightarrow$H$_j\rightarrow$O$_\mathrm{all}$ stage, if O$_i$ (selected in O$_\mathrm{all}$) is a neighbor of H$_j$, the derivative for O$_i$ coordinate is governed by the following equation \ref{si_eq:d_s_O_1}.
\begin{equation}
    \frac{\partial \mathbf{s(R)}}{\partial \mathbf{R}_{\mathrm{O}_i}} = \sum_{i \in \mathrm{O}} \sum_{j \in \mathrm{H}}-\lambda \frac{e^{-\lambda\left|\mathbf{R}_{\mathrm{O}_i}-\mathbf{R}_{\mathrm{H}_j}\right|}}{\sum_{m \in \mathrm{O}} e^{-\lambda\left|\mathbf{R}_{\mathrm{O}_m}-\mathbf{R}_{\mathrm{H}_j}\right|}}\left[1-\frac{e^{-\lambda\left|\mathbf{R}_{\mathrm{O}_i}-\mathbf{R}_{\mathrm{H}_j}\right|}}{\sum_{m \in \mathrm{O}} e^{-\lambda\left|\mathbf{R}_{\mathrm{O}_m}-\mathbf{R}_{\mathrm{H}_j}\right|}}\right] \frac{\mathbf{R}_{\mathrm{O}_i}-\mathbf{R}_{\mathrm{H}_j}}{\left|\mathbf{R}_{\mathrm{O}_i}-\mathbf{R}_{\mathrm{H}_j}\right|}
\label{si_eq:d_s_O_1}
\end{equation}
if O$_k$ (selected in O$_\mathrm{all}$, $k \neq i$) is a neighbor of H$_j$, the derivative for O$_k$ coordinate is given by the following equation \ref{si_eq:d_s_O_0}.
\begin{equation}
    \frac{\partial \mathbf{s(R)}}{\partial \mathbf{R}_{\mathrm{O}_k}} = \sum_{i \in \mathrm{O}} \sum_{j \in \mathrm{H}}-\lambda \frac{e^{-\lambda\left|\mathbf{R}_{\mathrm{O}_k}-\mathbf{R}_{\mathrm{H}_j}\right|}}{\sum_{m \in \mathrm{O}} e^{-\lambda\left|\mathbf{R}_{\mathrm{O}_m}-\mathbf{R}_{\mathrm{H}_j}\right|}}\left[0-\frac{e^{-\lambda\left|\mathbf{R}_{\mathrm{O}_i}-\mathbf{R}_{\mathrm{H}_j}\right|}}{\sum_{m \in \mathrm{O}} e^{-\lambda\left|\mathbf{R}_{\mathrm{O}_m}-\mathbf{R}_{\mathrm{H}_j}\right|}}\right] \frac{\mathbf{R}_{\mathrm{O}_k}-\mathbf{R}_{\mathrm{H}_j}}{\left|\mathbf{R}_{\mathrm{O}_k}-\mathbf{R}_{\mathrm{H}_j}\right|} \approx 0
\label{si_eq:d_s_O_0}
\end{equation}
At this point, all derivatives with respect to the coordinates have been determined. In addition, we incorporate the neighbor list function, which efficiently searches for nearby atoms, reducing the computational complexity from $\mathcal{O}(N[\mathrm{O_{all}} \times \mathrm{H_{all}} \times \mathrm{O_{all}}])$ to $\mathcal{O}(N[\mathrm{O_{all}} \times \mathrm{H_{neighbor}} \times \mathrm{O_{neighbor}}])$. In particular, a neighbor list with a cutoff $2.4$ {\AA} is set, updated at each step, to search for atoms in groups.

\section*{Polarization calculation}
In the realm of polarization analysis, a well-established formalism extensively relies on the utilization of maximally localized Wannier functions (MLWFs).\cite{marzari1997maximally,marzari2012maximally} In a molecular or condensed matter system, polarization arises due to the separation of positive and negative charges. It can be characterized by the dipole moment, which is a measure of the overall charge distribution within the system. When dealing with finite systems, the dipole $\boldsymbol{\mu}$ of the glycine molecule can be calculated as follows\cite{silvestrelli1999water,sharma2005intermolecular}:
\begin{equation}
\boldsymbol{\mu} = e\left(\sum_{i \in \mathrm{H}} \mathbf{R}_{i} + 4\sum_{j \in \mathrm{C}} \mathbf{R}_{j} + 5\sum_{k \in \mathrm{N}} \mathbf{R}_{k} + 6\sum_{l \in \mathrm{O}} \mathbf{R}_{l} - 2\sum_{m \in \mathrm{W}} \mathbf{R}_{m}\right)
\label{eq:dipole_mu}
\end{equation}
where $e$ represents the electronic charge. The positions of the nuclei (H, C, N, and O) and the MLWF centers (W) of glycine are considered in the calculation. Each nucleus without outer valence electrons (H, C, N, and O) carries a positive charge of $1$, $4$, $5$, and $6$, respectively. Additionally, each MLWF center is associated with $2$ valence electrons. The MLWF centers are determined through the calculation in CP2K\cite{kuhne2020cp2k}. The structures of solvated glycine are obtained from the enhanced sampling trajectory of system A (Table \ref{tab:system_box}). The remaining settings are kept consistent with the DFT labeling for the DeePKS dataset.

\section*{Other computational details}
\begin{figure}
  \centering
  \includegraphics[width=\textwidth]{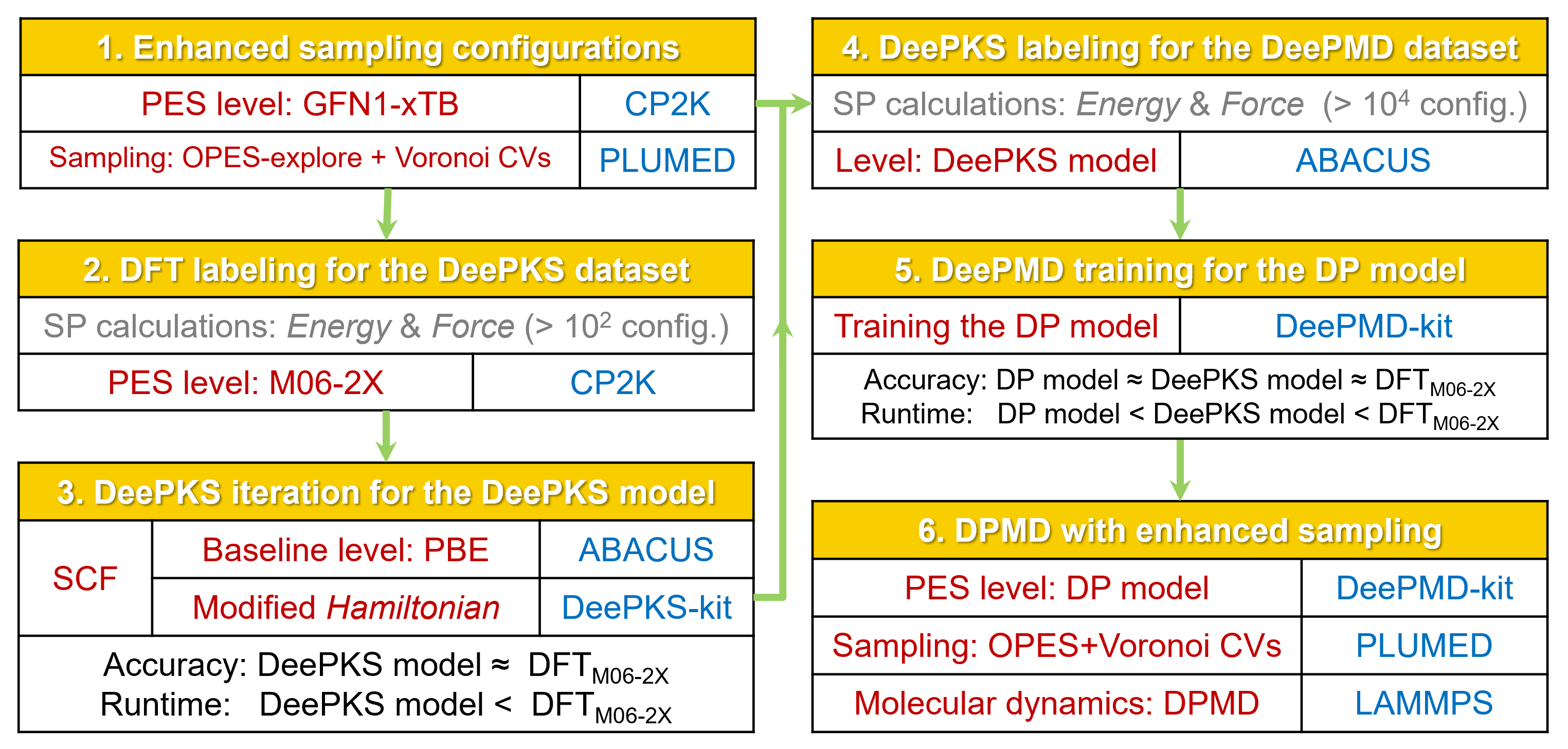}
  \caption{\textbf{Simulation workflow in detail.} The comprehensive workflow includes enhanced sampling, DFT labeling, DeePKS iteration, DeePKS labeling, DP training, and DPMD with OPES for efficient and accurate exploration of glycine tautomerism in water.}
  \label{si_fig_workflow}
\end{figure}

Hydrogen bond (HB) analysis is performed using the MDAnalysis\cite{michaud2011mdanalysis,gowers2016mdanalysis} library, where the distance cutoff between the donor and the acceptor is $3.5$ {\AA}, and the donor-hydrogen-acceptor angle cutoff is set to $140 \degree$. To compare the explicit solvation effect, the glycine clusters are also modeled using the implicit solvation model (SMD)\cite{marenich2009universal} at the M06-2X/def2-TZVP level\cite{zhao2008m06,weigend2005balanced,weigend2006accurate} in the Gaussian \cite{g16}. The simulation results are visualized using the VMD\cite{humphrey1996vmd} and Matplotlib\cite{hunter2007matplotlib} packages. The diffusion coefficient is calculated from the mean square displacement (MSD) using Einstein's relation. The diffusion coefficient of water is $2.36\pm0.09\times10^{-9}$ m$^2$/s at $300$ K, which is consistent with experiments (\textasciitilde $ 2.3\times10^{-9}$ m$^2$/s). The general simulation workflow (Fig. \ref{si_fig_workflow}) comprises six steps mentioned in manuscript. 

\section*{Solvation and polarization}

Notably, the ongoing research into the minimum number of water molecules required to stabilize the [Z] form has remained controversial, with debates ranging from two to ten water molecules, and means that effective treatment and accurate description of the solvent effect remains a crucial and challenging task.\cite{perez2016water,ramaekers2004neutral,aikens2006incremental,bachrach2008microsolvation,jensen1995number,tripathi2021unveiling,kaufmann2016helium,kim2014gas}. We discuss the solvation and polarization of glycine in the SI, which in turn fine-tune the thermodynamics and dynamic behavior.\cite{azizi2023solvation}

\begin{figure}
  \centering
  \includegraphics[width=\textwidth]{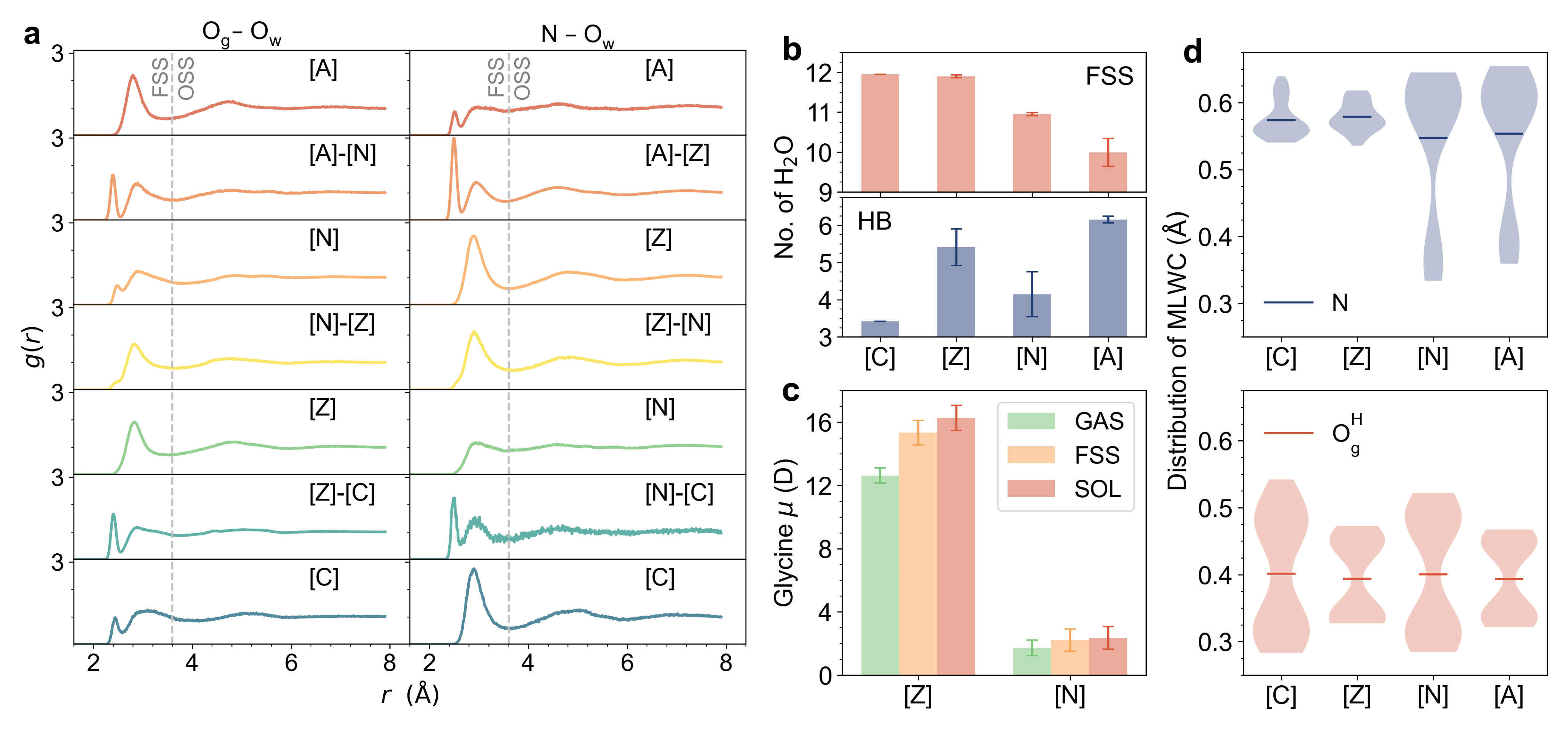}
  \caption{\textbf{The solvation and polarization of glycine.} 
  \textbf{a}, RDF profiles with respect to glycine transformations for the pairs of particles comprising $\mathrm{O_g}$-$\mathrm{O_w}$ and N-$\mathrm{O_w}$ targeted atoms. The region within a specified cutoff distance of 3.6 {\AA} is defined as the FSS of glycine and the external region is described as the OSS.
  \textbf{b}, The number of water molecules and total HB number of glycine in the FSS. 
  \textbf{c}, The change of the glycine dipole moment in gas phase (GAS), FSS, and full solvent (SOL), where the molecular structures of glycine in gas phase and FSS are those obtained from the fully solvated glycine by removing corresponding water molecules. 
  \textbf{d}, The violin plots represent the distributions of the distances between the nuclei (N and $\mathrm{O_g^H}$ ) of fully solvated glycine and the MLWCs, where the lines represent the mean value.}
  \label{fig_RDF_HB_Dip}
\end{figure}

The gradual expansion of the solvent shell is studied by the radial distribution function (RDF)\cite{wood2008nature,sun2010glycine} to understand the structural distribution between glycine and water molecules, as depicted in the Fig. \ref{fig_RDF_HB_Dip}a. In the region of the first solvent shell (FSS), the RDFs of $\mathrm{O_g-O_w}$ display a peak shift between 2.4 and 2.8 {\AA} in the [A]$\leftrightarrow$[N] or [Z]$\leftrightarrow$[C] path, which is relative to the protonation of the --COO$^-$ group. The situation is similar to the protonation of the --NH$_2$ group in the  [A]$\leftrightarrow$[Z] or [N]$\leftrightarrow$[C] path, where the RDFs of $\mathrm{N-O_w}$ have a peak shift between 2.5 and 2.9 {\AA}. In the outer solvent shell (OSS), RDFs show slight changes between 3.6 {\AA} and 6 {\AA} depending on changes in FSS. Minimal variations are observed beyond the homogeneous region that extends beyond 6 {\AA}, suggesting that FSS predominantly influences the solvation of glycine, and this is also supported by the THz spectra.\cite{sun2014understanding}

In the FSS, the [Z], [N], [A], and [C] forms are associated with approximately 12, 11, 10, and 12 water molecules, and exhibit 5.4, 4.1, 6.2, and 3.4 HBs, respectively (Fig. \ref{fig_RDF_HB_Dip}b). Compared to other forms, the stability of the [Z] form can be attributed to the greater abundance of water molecules within the FSS and the formation of a more extensive HB network, which also contribute to the polarization of [Z]. 

In full solution, the [Z] form exhibits an average dipole moment of $16.3\pm0.8$ D, whereas in its isolated form, the dipole moment amounts to only $12.6\pm0.5$ D (Fig. \ref{fig_RDF_HB_Dip}c). These results align with previous calculation\cite{sun2010glycine}. Remarkably, when exclusively considering the FSS, a dipole moment of $15.4\pm0.7$ D is emulated, effectively mirroring the characteristic polarization of the [Z] form. The [N] form, whether in a gaseous phase or within an aqueous environment, exhibits dipole moments typically ranging from 1.6-2.3 D. This is attributed to the non-ionized state of both the amino and carboxyl groups in the [N] form. In contrast, the presence of water molecules in the FSS mainly stabilizes the charged --COO$^-$ and --NH$_3$$^+$ groups in the [Z] form, leading to an augmentation of the overall dipole moment. 

Specifically, the average distribution value of the electron pair (represented by maximally localized Wannier centers (MLWCs)\cite{marzari1997maximally,marzari2012maximally}) surrounding $\mathrm{O_g^H}$ is approximately 0.15 {\AA} smaller than that of N, and the electron pairs near $\mathrm{O_g^H}$ exhibit greater localization compared to that near N (Fig. \ref{fig_RDF_HB_Dip}d). This indicates that --COO$^-$ primarily influences the larger dipole moment as opposed to --NH$_3$$^+$, due to the higher electronegativity of $\mathrm{O_g^H}$.

\newpage
\begin{table}[!ht]
        \renewcommand{\arraystretch}{1.2}
	\caption{\textbf{Details of the simulation systems.} systems A and B are used for building dataset, as well as system C are used for conducting molecular dynamics simulations and subsequent data analysis.}
	\centering
	\begin{tabular}{ccccc}
		\toprule
            System & Usage & Box size (Å\textsuperscript{3}) & No. of H\textsubscript{2}O & No. (Molar) of glycine\\
		\midrule
            A & Training, testing dataset & 12 $\times$ 12 $\times$ 12 & 54 & 1 (1.03 M)\\
            B & Training, testing dataset   & 12 $\times$ 12 $\times$ 12       & 58 & 0 \\
            C & Final MD and analysis & 15.8 $\times$ 15.8 $\times$ 15.8 & 128 & 1 (0.43 M)\\		
            \bottomrule
	\end{tabular}
	\label{tab:system_box}
\end{table}

\begin{table}[!ht]
        \renewcommand{\arraystretch}{1.2}
	\caption{\textbf{A convergence test of the single point calculation.} The plane-wave cutoff test of DFT labeling for the DeePKS dataset. The calculation of $\Delta E$ involves subtracting previous energy in order to determine the difference, for example, $\Delta E(700 \mathrm{Ry},60 \mathrm{Ry}) = E(700 \mathrm{Ry},60 \mathrm{Ry})-E(600 \mathrm{Ry},60 \mathrm{Ry})$, $\Delta E(800 \mathrm{Ry},60 \mathrm{Ry}) = E(800 \mathrm{Ry},60 \mathrm{Ry})-E(700 \mathrm{Ry},60 \mathrm{Ry})$ and so on.}
	\centering
	\begin{tabular}{ccccc}
		\toprule
            Cutoff (Ry) & Rel\_Cutoff (Ry) & $\Delta E$ (a.u.) & $\Delta E$ (a.u./atom) \\
		\midrule
            600  & 60  & Basis$_1$ & Basis$_1$/atom \\
            700  & 60  & $-$3.23 $\times$ 10\textsuperscript{-3}  & $-$1.88 $\times$ 10\textsuperscript{-5} \\
            800  & 60  &  5.44 $\times$ 10\textsuperscript{-4}  &  3.16 $\times$ 10\textsuperscript{-6} \\
            900  & 60  & $-$5.89 $\times$ 10\textsuperscript{-5}  & $-$3.43 $\times$ 10\textsuperscript{-7} \\
            1000 & 60  & $-$9.65 $\times$ 10\textsuperscript{-5}  & $-$5.61 $\times$ 10\textsuperscript{-7} \\
            1100 & 60  &  8.75 $\times$ 10\textsuperscript{-8}  &  5.09 $\times$ 10\textsuperscript{-10} \\
            \cline{1-4} 
            1000 & 60  & Basis$_2$ & Basis$_2$/atom \\
            1000 & 70  &  1.29 $\times$ 10\textsuperscript{-7}  &  7.51 $\times$ 10\textsuperscript{-10} \\
            1000 & 80  & $-$2.20 $\times$ 10\textsuperscript{-9}  & $-$1.28 $\times$ 10\textsuperscript{-11} \\
            1000 & 90  &  4.00 $\times$ 10\textsuperscript{-10} &  2.33 $\times$ 10\textsuperscript{-12} \\
            1000 & 100 &  1.00 $\times$ 10\textsuperscript{-10} &  5.82 $\times$ 10\textsuperscript{-13} \\
            1000 & 110 &  3.00 $\times$ 10\textsuperscript{-10} &  1.74 $\times$ 10\textsuperscript{-12} \\		
            \bottomrule
	\end{tabular}
	\label{tab:planewave_cutoff}
\end{table}

\begin{table}[!ht]
        \renewcommand{\arraystretch}{1.2}
	\caption{\textbf{The runtime of calculations.} DFT labelling and DeePKS labelling are performed on CPU processor with parallelization, where the runtime is for a configuration to complete a self-consistent field iteration. The runtime of DPMD simulation with OPES on GPU is also given.}
	\centering
	\begin{tabular}{cccc}
		\toprule
            Step & System & Machine & Time \\
		\midrule
            DFT labeling for the DeePKS dataset & A & 24-core CPU  & 40 min/frame \\
            DFT labeling for the DeePKS dataset & B & 24-core CPU  & 35 min/frame\\
            DeePKS labeling for the DP dataset  & A & 24-core CPU  & 5 min/frame\\
            DeePKS labeling for the DP dataset  & B & 24-core CPU  & 4 min/frame \\
            DPMD simulation with OPES                   & A & 1 Tesla V100 & 70 timesteps/s \\
            DPMD simulation with OPES                   & C & 1 Tesla V100 & 12 timesteps/s \\
            \bottomrule
	\end{tabular}
	\label{tab:scf_time}
\end{table}

\clearpage
\begin{table}[!ht]
        \renewcommand{\arraystretch}{1.2}
	\caption{\textbf{The component of the datasets and the accuracy of the models.} The number of configurations in the train and testing datasets for systems A and B (the DeePKS training datasets and all test datesets are performed by M06-2X/QZV3P, and the DP training datasets are performed by the DeePKS models). The root mean square errors (RMSEs) for the DeePKS and DP models on the test data sets are also given.}
	\centering
	\begin{tabular}{ccccc}
		\toprule
            Infomation & DeePKS (syst. A) & DeePKS (syst. B) & DP (syst. A) & DP (syst. B) \\
		\midrule
            training dataset & 150 & 150 & 42,248 & 13,250 \\
            testing dataset & 205 & 165 & 5,560 & 460 \\
            $E_\mathrm{RMSE}$ (meV/atom) & 0.53 & 0.61 & 0.79 & 1.72 \\
            $F_\mathrm{RMSE}$ (meV/Å) & 43 & 52 & 58 & 63 \\		
            \bottomrule
	\end{tabular}
	\label{tab:model_accuracy}
\end{table}

\clearpage

\begin{figure}
  \centering
  \includegraphics[width=\textwidth]{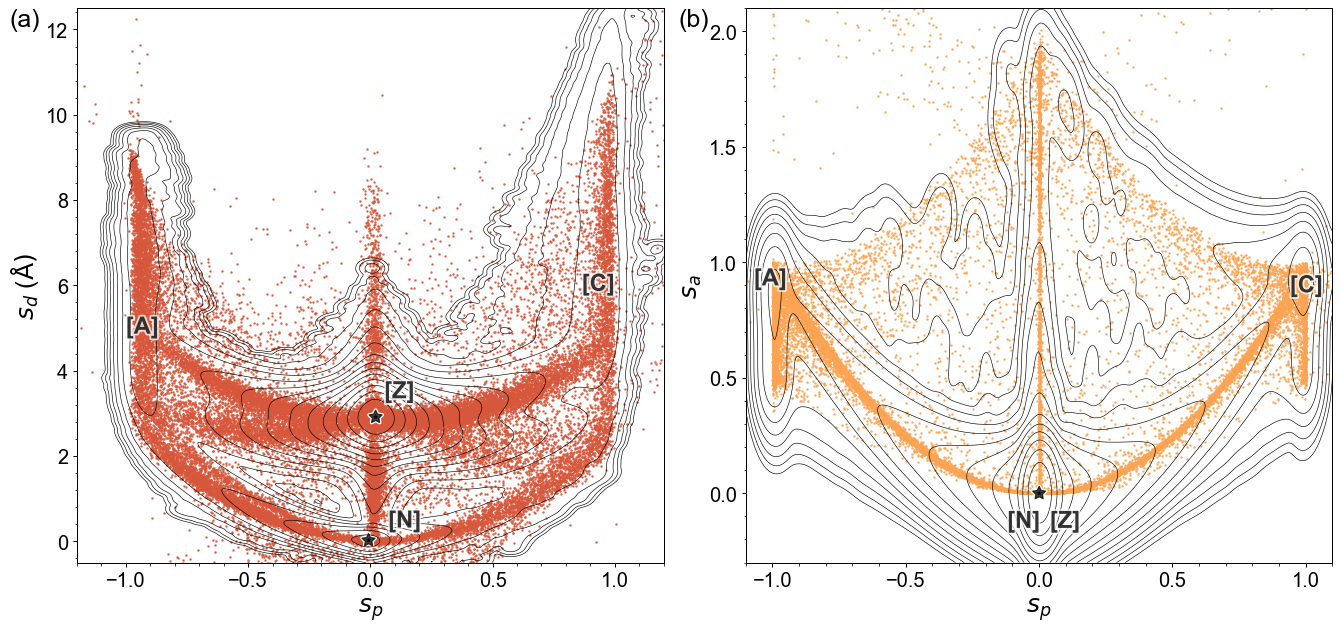}
  \caption{\textbf{The distribution of the configuration in the training dataset.} The configurational distribution with respect to (a) glycine protonation ($\mathbf{s}_p^{\lambda=5}$) and charge-charge distance ($\mathbf{s}_d^{\lambda=5}$), (b) glycine protonation ($\mathbf{s}_p^{\lambda=8}$) and number of self-ions ($\mathbf{s}_a^{\lambda=8}$). The origin of Contours is attributed to the utilization of free energy reweighting. The training dataset contains sufficient coverage of a satisfactory phase space using the Voronoi CVs.}
  \label{si_fig_train_dataset_cv_gly}
\end{figure}

\begin{figure}
  \centering
  \includegraphics[width=0.75\textwidth]{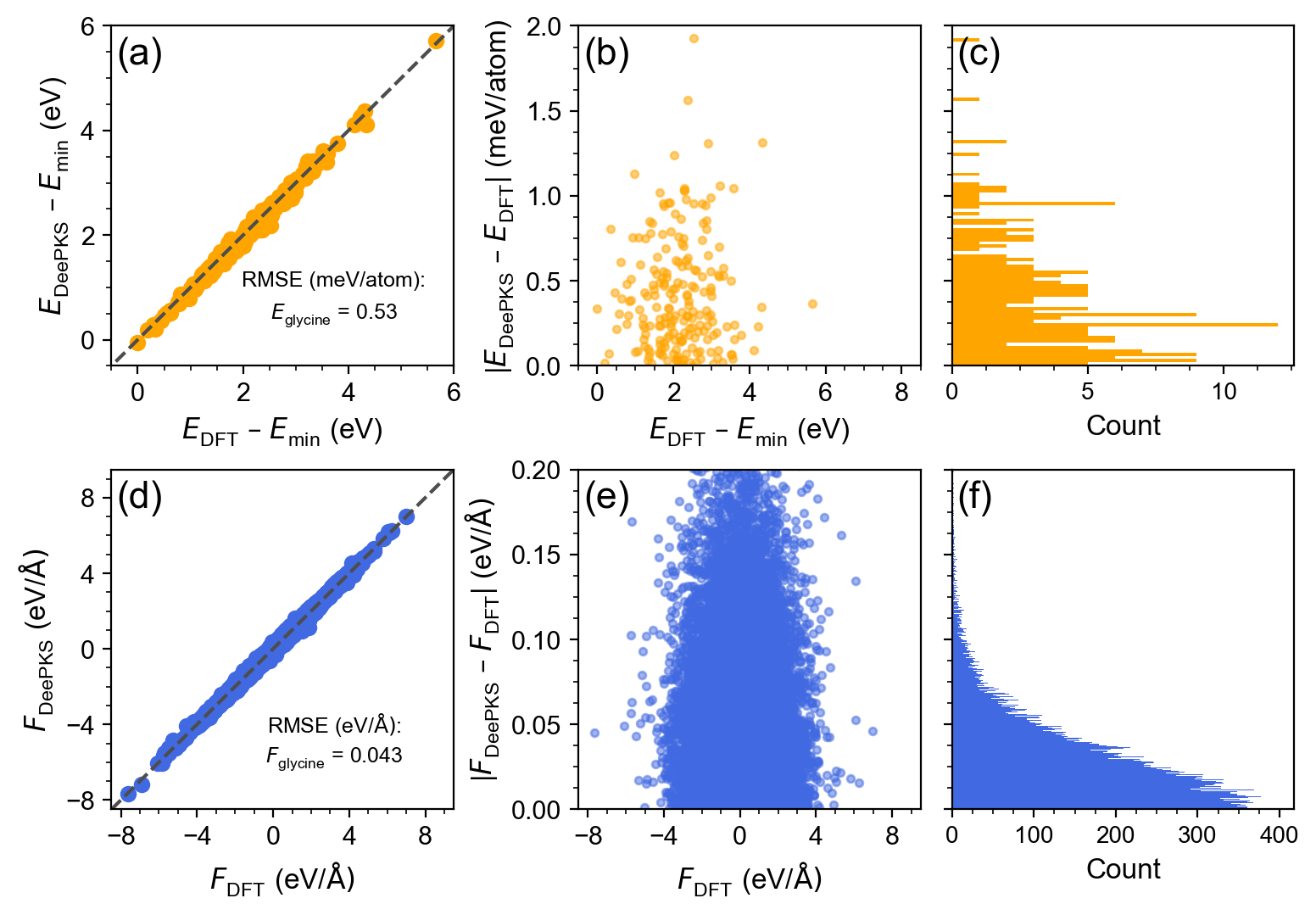}
  \caption{\textbf{Error distribution of the DeePKS model on the testing dataset of glycine in water (system A).} (a) The distribution of $E_{\mathrm{DeePKS}}-E_{\mathrm{min}}$ with respect to $E_{\mathrm{DFT}}-E_{\mathrm{min}}$. (b) Left: the distribution of $| E_{\mathrm{DeePKS}}-E_{\mathrm{DFT}} |$ with respect to $E_{\mathrm{DFT}}-E_{\mathrm{min}}$. (c) Right: the histogram of $| E_{\mathrm{DeePKS}}-E_{\mathrm{DFT}} |$. (d) The distribution of $F_{\mathrm{DeePKS}}$ with respect to $F_{\mathrm{DFT}}$. (e) Left: the distribution of $| F_{\mathrm{DeePKS}}-F_{\mathrm{DFT}} |$ with respect to $F_{\mathrm{DFT}}$. (f) Right: the histogram of $| F_{\mathrm{DeePKS}}-F_{\mathrm{DFT}} |$. Here, $E_{\mathrm{DFT}}$ and $F_{\mathrm{DFT}}$ are the energy and force calculated by the M06-2X functional; $E_{\mathrm{DeePKS}}$ and $F_{\mathrm{DeePKS}}$ are the energy and force performed by the present DeePKS model; and $E_{\mathrm{min}}$ is the minimum absolute energy in the testing dataset.}
  \label{si_fig_gly_dpks_test}
\end{figure}

\begin{figure}
  \centering
  \includegraphics[width=0.75\textwidth]{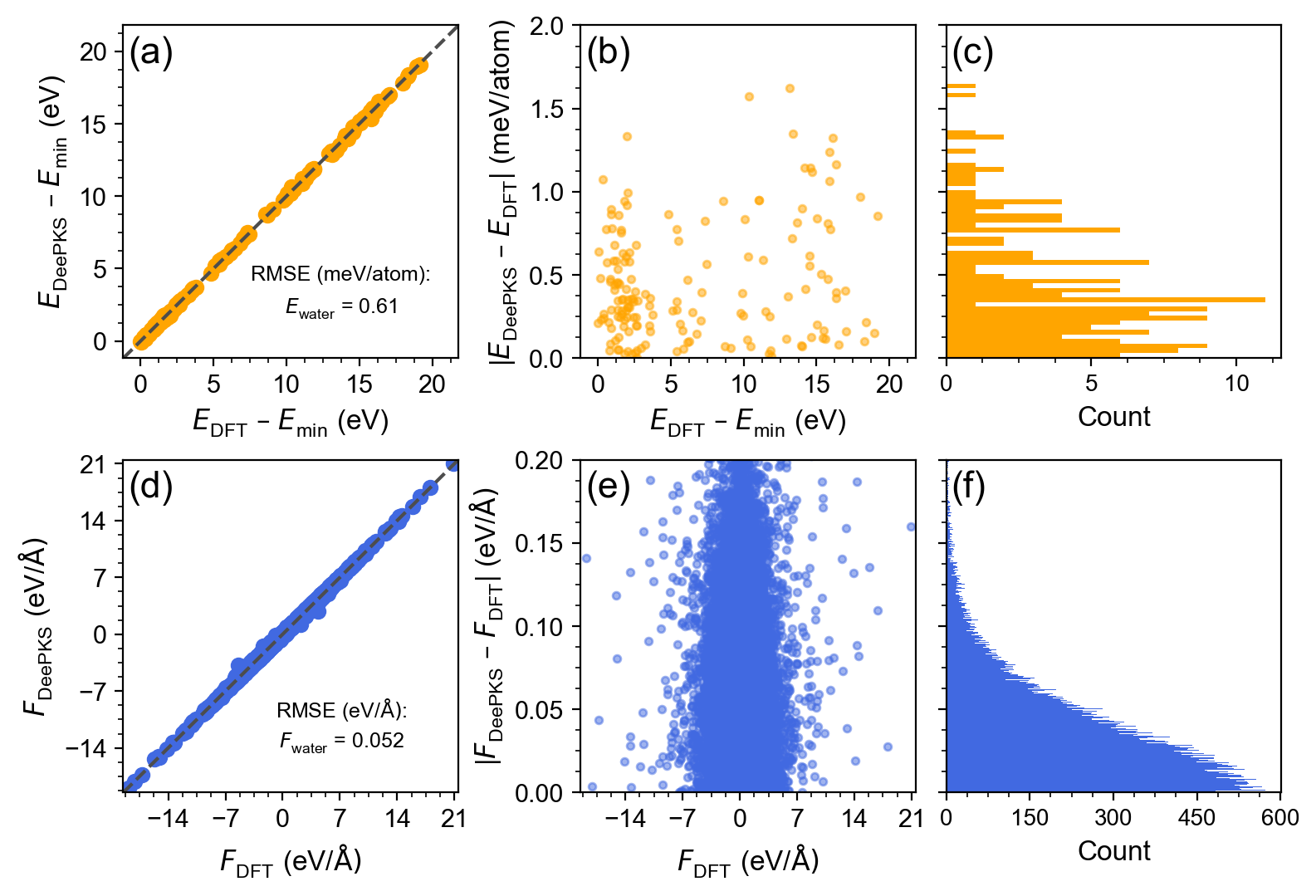}
  \caption{\textbf{Error distribution of the DeePKS model on the testing dataset of water (system B).} (a) The distribution of $E_{\mathrm{DeePKS}}-E_{\mathrm{min}}$ with respect to $E_{\mathrm{DFT}}-E_{\mathrm{min}}$. (b) Left: the distribution of $| E_{\mathrm{DeePKS}}-E_{\mathrm{DFT}} |$ with respect to $E_{\mathrm{DFT}}-E_{\mathrm{min}}$. (c) Right: the histogram of $| E_{\mathrm{DeePKS}}-E_{\mathrm{DFT}} |$. (d) The distribution of $F_{\mathrm{DeePKS}}$ with respect to $F_{\mathrm{DFT}}$. (e) Left: the distribution of $| F_{\mathrm{DeePKS}}-F_{\mathrm{DFT}} |$ with respect to $F_{\mathrm{DFT}}$. (f) Right: the histogram of $| F_{\mathrm{DeePKS}}-F_{\mathrm{DFT}} |$.}
  \label{si_fig_h2o_dpks_test}
\end{figure}

\begin{figure}
  \centering
  \includegraphics[width=0.75\textwidth]{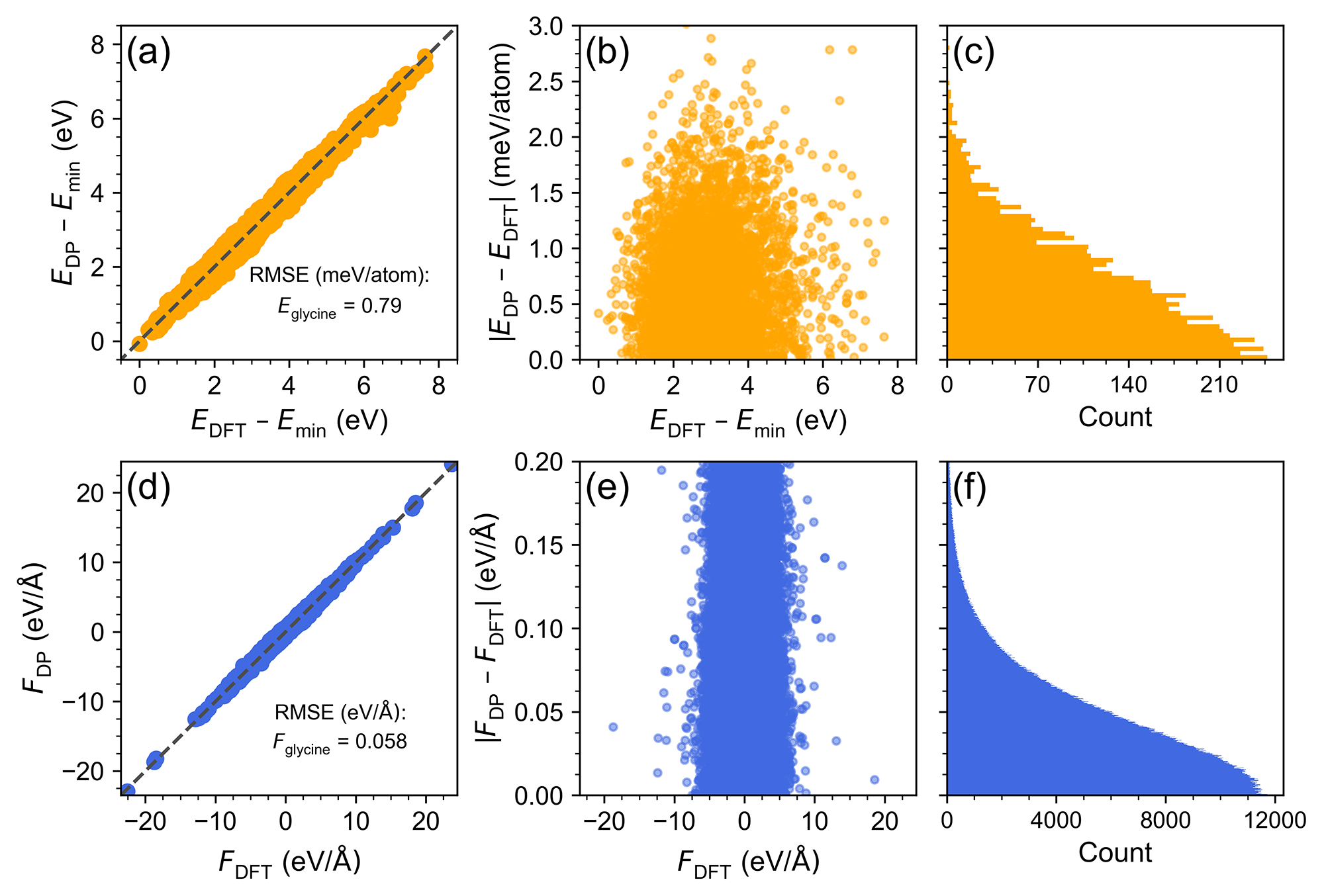}
  \caption{\textbf{Error distribution of the DP model on the testing dataset of glycine in water (system A).} (a) The distribution of $E_{\mathrm{DP}}-E_{\mathrm{min}}$ with respect to $E_{\mathrm{DFT}}-E_{\mathrm{min}}$. (b) Left: the distribution of $| E_{\mathrm{DP}}-E_{\mathrm{DFT}} |$ with respect to $E_{\mathrm{DFT}}-E_{\mathrm{min}}$. (c) Right: the histogram of $| E_{\mathrm{DP}}-E_{\mathrm{DFT}} |$. (d) The distribution of $F_{\mathrm{DP}}$ with respect to $F_{\mathrm{DFT}}$. (e) Left: the distribution of $| F_{\mathrm{DP}}-F_{\mathrm{DFT}} |$ with respect to $F_{\mathrm{DFT}}$. (f) Right: the histogram of $| F_{\mathrm{DP}}-F_{\mathrm{DFT}} |$. Here, $E_{\mathrm{DFT}}$ and $F_{\mathrm{DFT}}$ are the energy and force calculated by the M06-2X functional; $E_{\mathrm{DP}}$ and $F_{\mathrm{DP}}$ are the energy and force performed by the present DP model; and $E_{\mathrm{min}}$ is the minimum absolute energy in the testing dataset.}
  \label{si_fig_gly_dpmd_test}
\end{figure}

\begin{figure}
  \centering
  \includegraphics[width=0.75\textwidth]{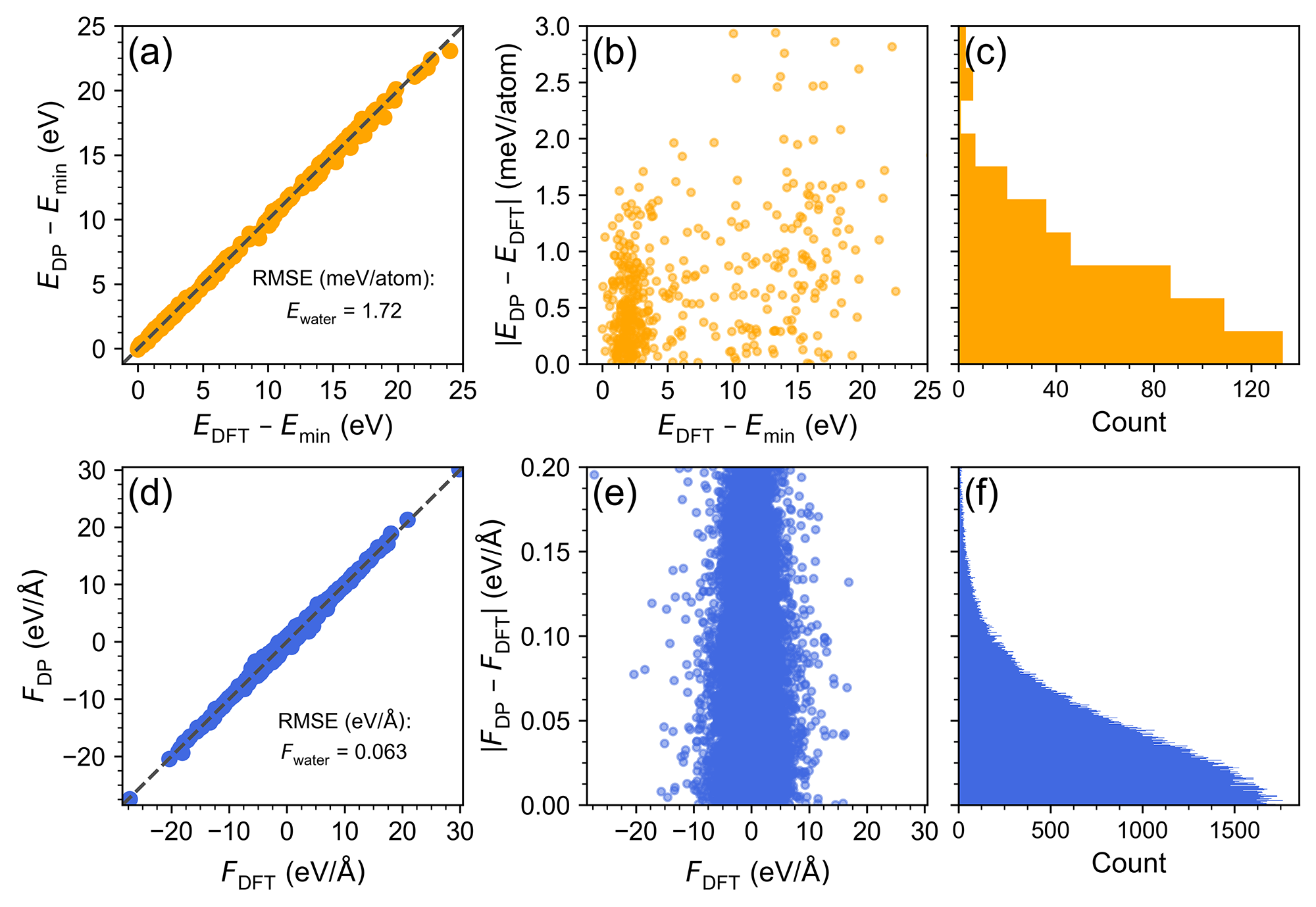}
  \caption{\textbf{Error distribution of the DP model on the testing dataset of water (system B).} (a) The distribution of $E_{\mathrm{DP}}-E_{\mathrm{min}}$ with respect to $E_{\mathrm{DFT}}-E_{\mathrm{min}}$. (b) Left: the distribution of $| E_{\mathrm{DP}}-E_{\mathrm{DFT}} |$ with respect to $E_{\mathrm{DFT}}-E_{\mathrm{min}}$. (c) Right: the histogram of $| E_{\mathrm{DP}}-E_{\mathrm{DFT}} |$. (d) The distribution of $F_{\mathrm{DP}}$ with respect to $F_{\mathrm{DFT}}$. (e) Left: the distribution of $| F_{\mathrm{DP}}-F_{\mathrm{DFT}} |$ with respect to $F_{\mathrm{DFT}}$. (f) Right: the histogram of $| F_{\mathrm{DP}}-F_{\mathrm{DFT}} |$.}
  \label{si_fig_h2o_dpmd_test}
\end{figure}

\begin{figure}
  \centering
  \includegraphics[width=\textwidth]{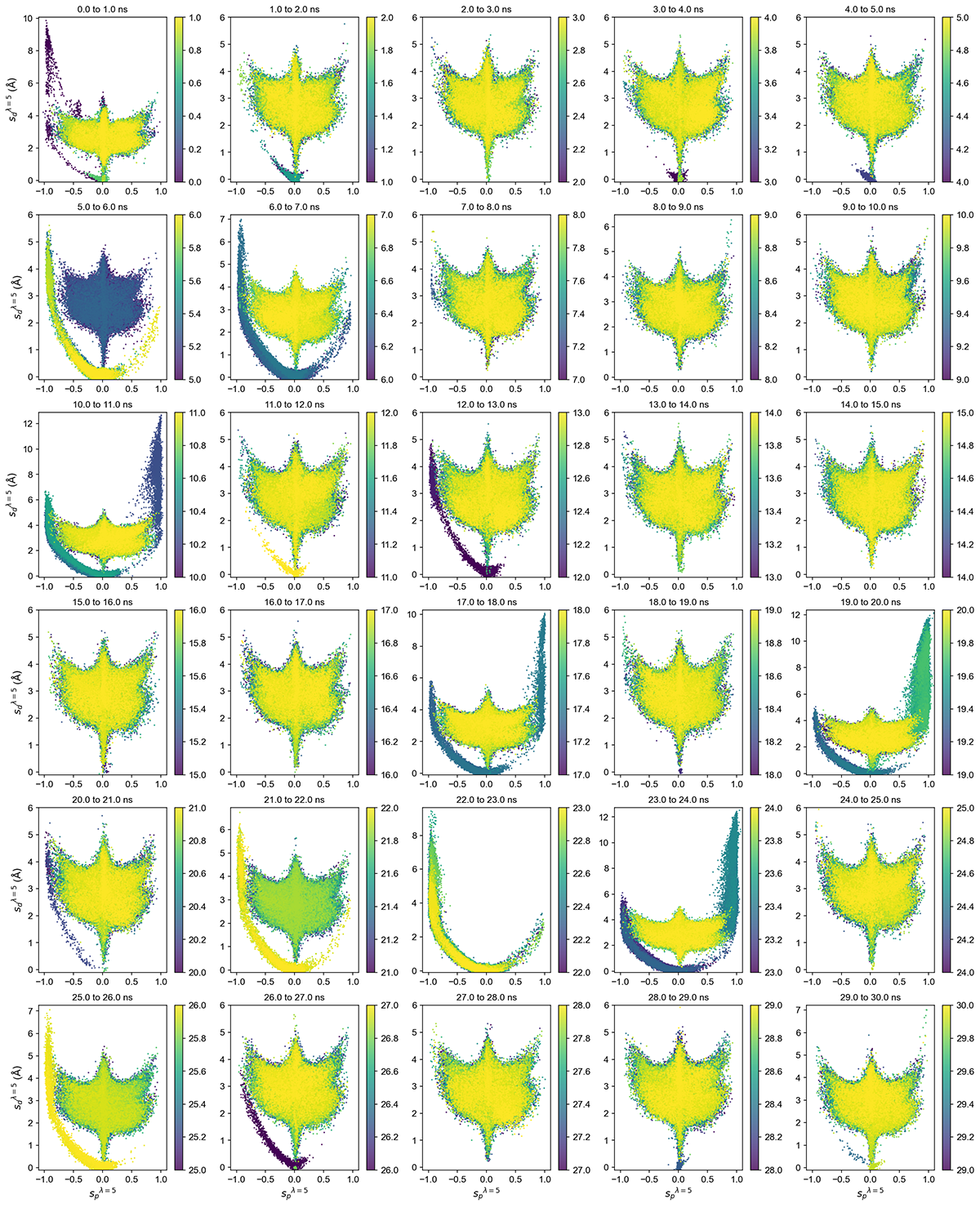}
  \caption{\textbf{The CV profiles with time.} The evolution of glycine protonation ($\mathbf{s}_p^{\lambda=5}$) and charge-charge distance ($\mathbf{s}_d^{\lambda=5}$), with each configuration assigned a color code according to the corresponding simulation time represented in the color bar. The sampling shows a discernible tendency to resemble the 'Maple Leaf' and 'U' shapes, and is multiple traversals of all possible configurations.}
  \label{si_fig_sp_sd_time}
\end{figure}

\begin{figure}
  \centering
  \includegraphics[width=\textwidth]{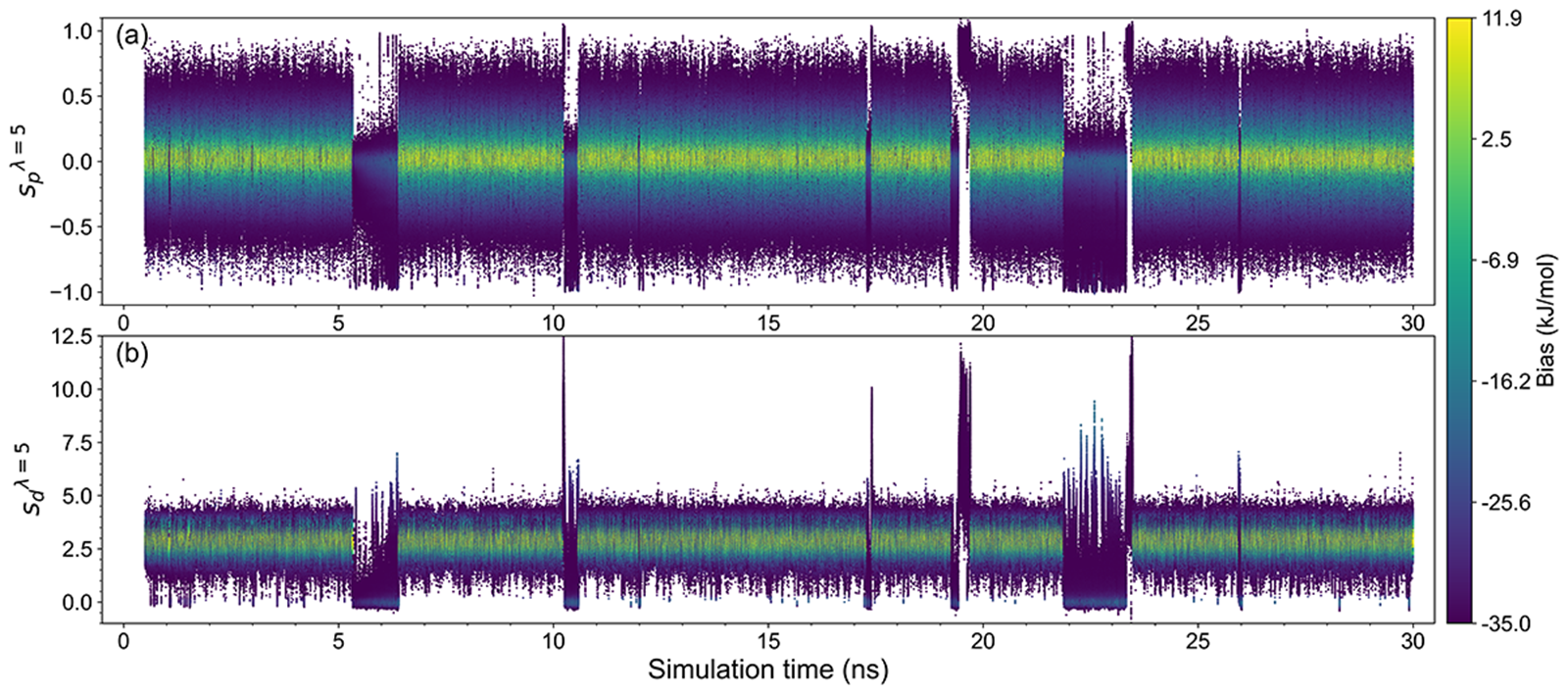}
  \caption{\textbf{The CV profiles with bias.} The evolution of (a) glycine protonation ($\mathbf{s}_p^{\lambda=5}$) and (b) charge-charge distance ($\mathbf{s}_d^{\lambda=5}$) over simulation time, with each configuration assigned a color code according to the corresponding bias value represented in the color bar. Based on extensive enhanced sampling, the bias approximately reaches a quasi-static state.}
  \label{si_fig_sp_sd_bias}
\end{figure}

\begin{figure}
  \centering
  \includegraphics[width=\textwidth]{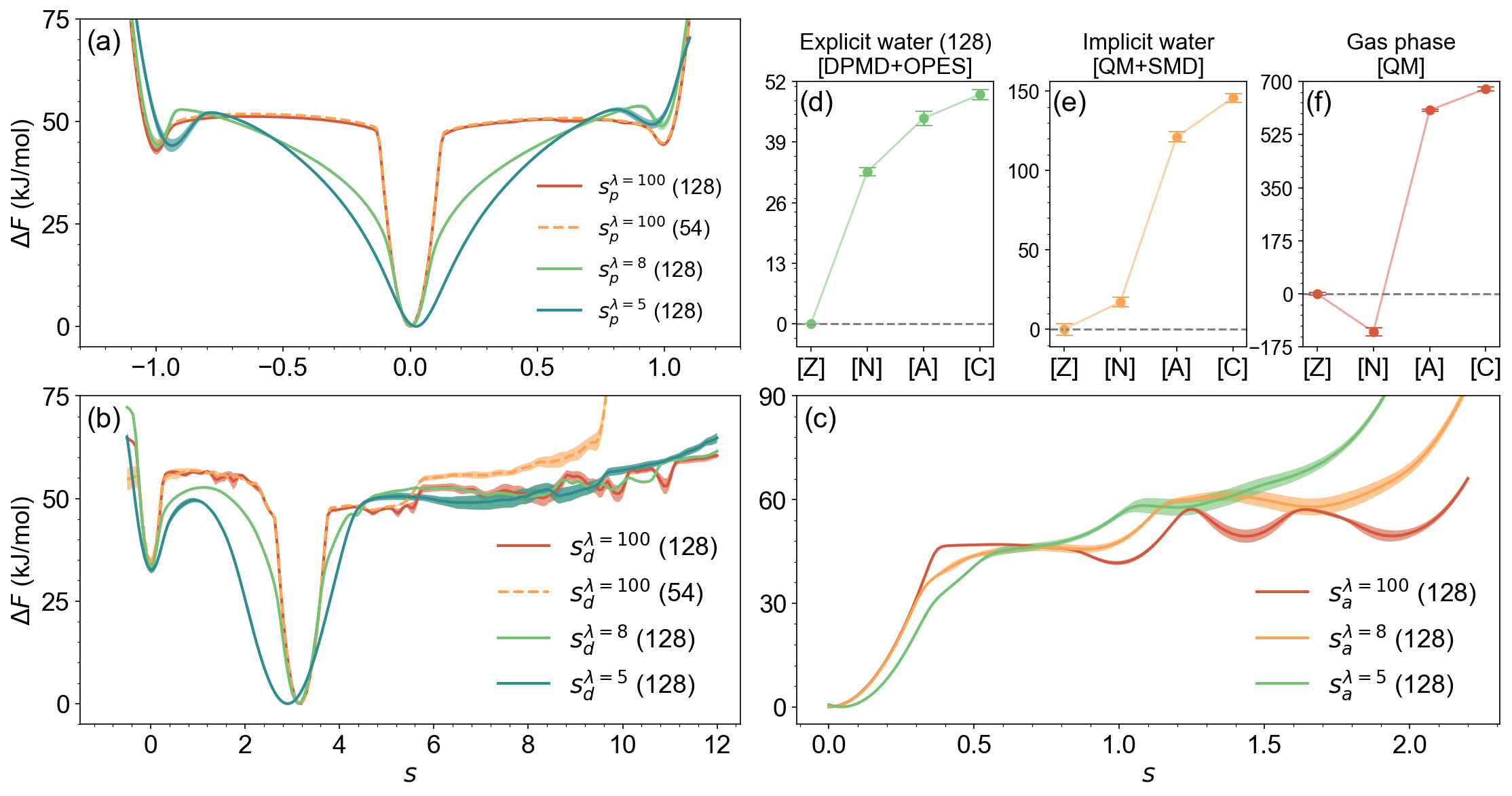}
  \caption{\textbf{The one-dimensional free energy and free energy differences of different glycine forms.} The free energy profiles with respect to CVs (a) $\mathbf{s}_p$, (b) $\mathbf{s}_d$, and (c) $\mathbf{s}_a$, where 128 and 54 represent the number of $\mathrm{H_2O}$ with a glycine molecule. The inclusion of 128 $\mathrm{H_2O}$ takes into account sufficient size effects and enables converged free energy compared to 54 $\mathrm{H_2O}$. The free energy differences (kJ/mol) among glycine in [Z], [N], [A], and [C] forms under (d) explicit water solvation as simulated in this work [DPMD+OPES], (e) implicit water model [QM+SMD], and (f) gas phase [QM]. If we consider the implicit solvation of glycine, only quantitative results can be obtained, suggesting that the interactions ignored by implicit solvation models influence the stabilization of glycine. The QM calculation employed the chemical model method of M06-2X/def2-TZVP. }
  \label{si_fig_1d_fes}
\end{figure}

\begin{figure}
  \centering
  \includegraphics[width=0.6\textwidth]{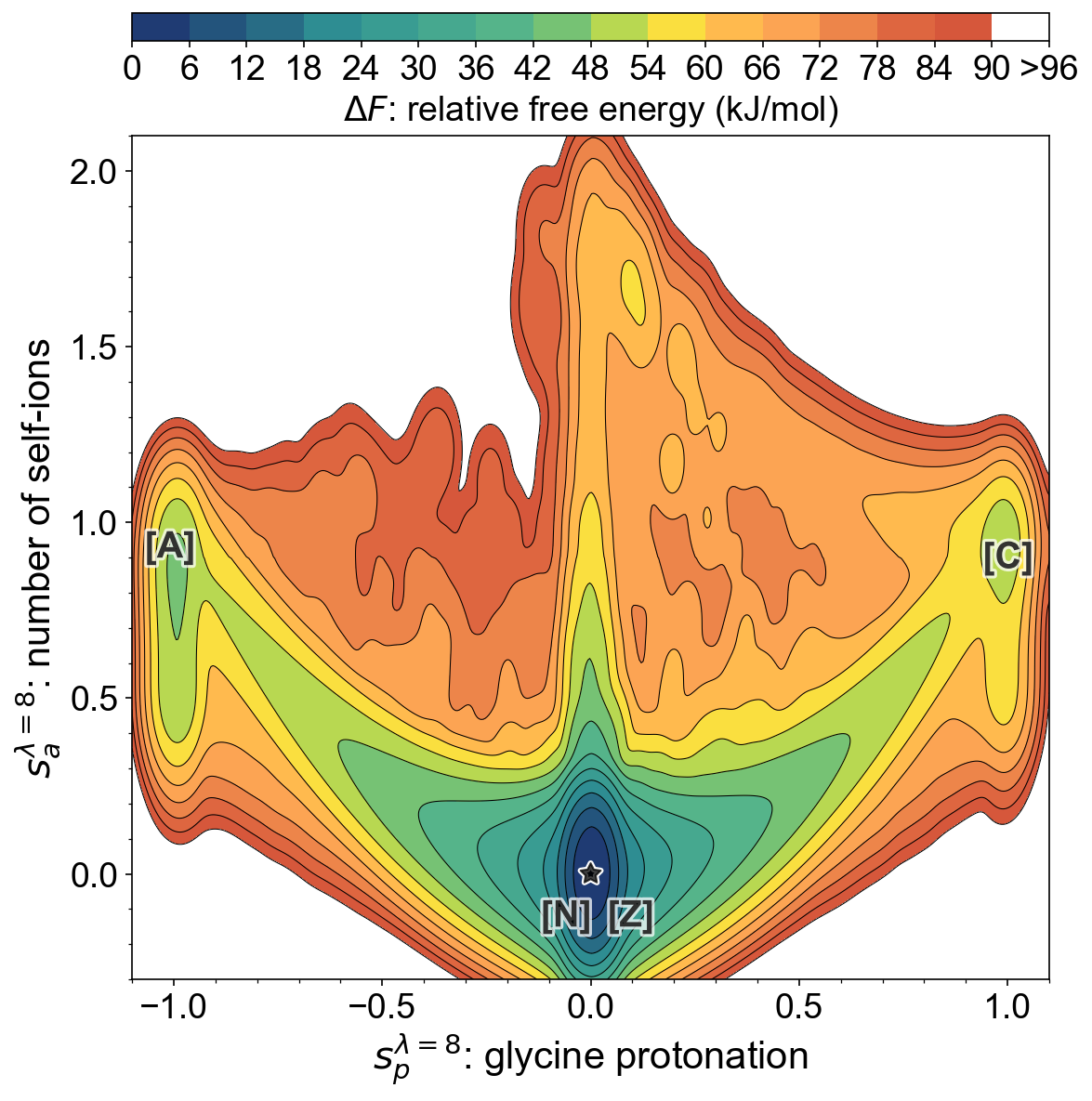}
  \caption{\textbf{Two-dimensional free energy surface (FES).} FES with respect to glycine protonation ($\mathbf{s}_p^{\lambda=8}$) and number of self-ions ($\mathbf{s}_a^{\lambda=8}$). There is a H$_3$O$^+$ with [A] states ($\mathbf{s}_a^{\lambda=8} \approx 1$), an OH$^-$ with [C] state ($\mathbf{s}_a^{\lambda=8} \approx 1$), and no ion in [N] or [Z] state ($\mathbf{s}_a^{\lambda=8} \approx 0$).}
  \label{si_fig_2d_fes_sc}
\end{figure}

\begin{figure}
  \centering
  \includegraphics[width=\textwidth]{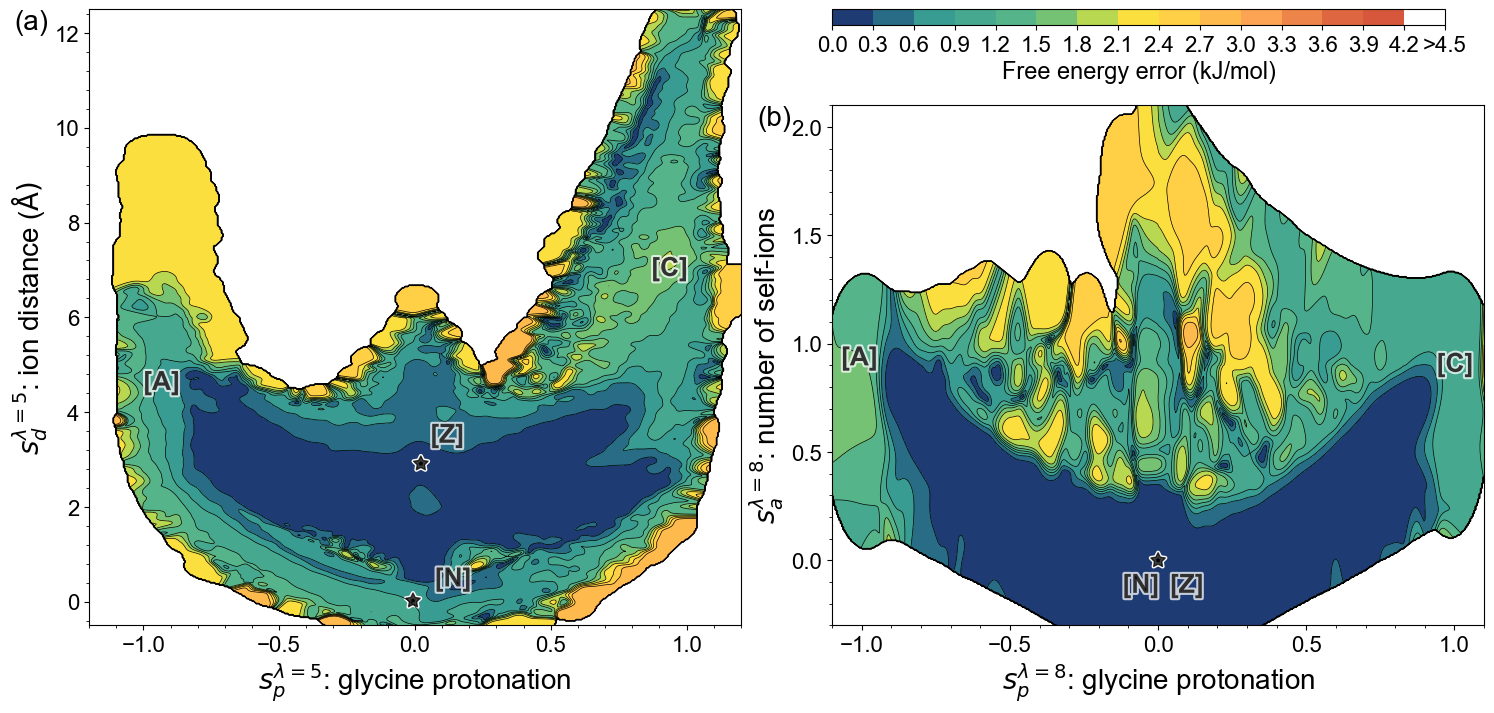}
  \caption{\textbf{The standard deviation of free energy surfaces}. Two-dimensional (b) free energy error with respect to glycine protonation ($\mathbf{s}_p^{\lambda=5}$) and charge-charge distance ($\mathbf{s}_d^{\lambda=5}$), and (c) free energy error with respect to glycine protonation ($\mathbf{s}_p^{\lambda=8}$) and number of self-ions ($\mathbf{s}_a^{\lambda=8}$).}
  \label{si_fig_err_fes}
\end{figure}

\begin{figure}
  \centering
  \includegraphics[width=0.6\textwidth]{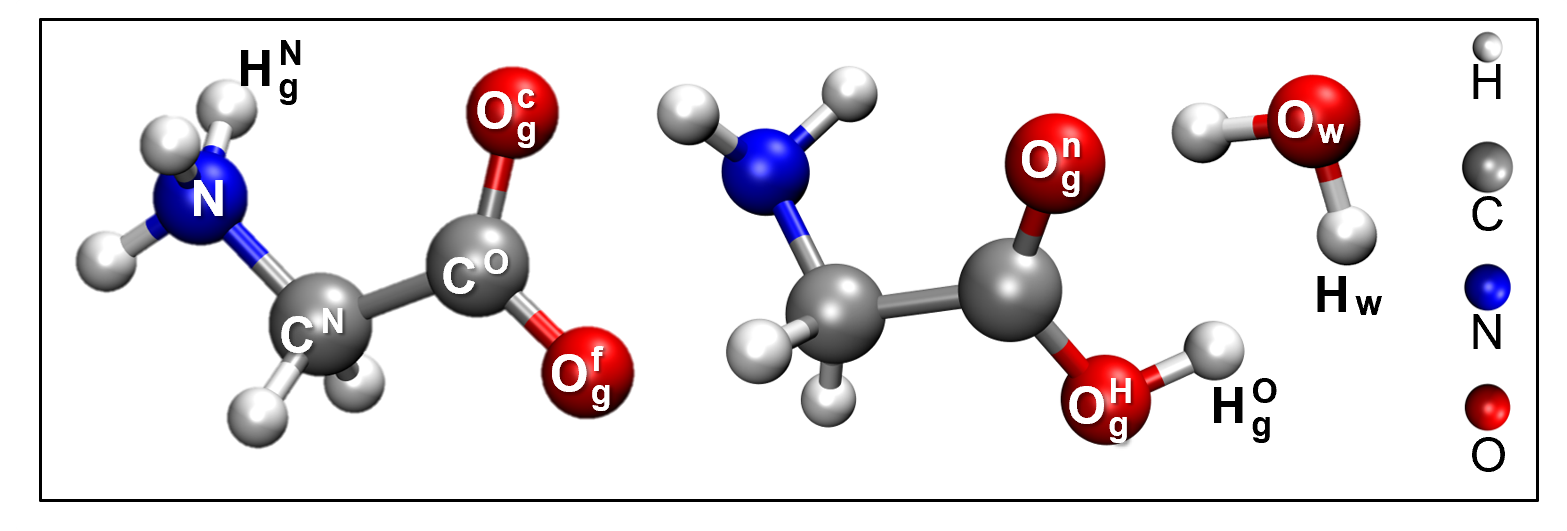}
  \caption{\textbf{Atom notations} are used to denote the atoms present in the molecules of glycine and water. $\mathrm{X_y}$ represents the atom $\mathrm{X}$ (H or O) that originates from molecule y (glycine or water). Specifically, for the --COO$^-$ group, $\mathrm{O^c}$ ($\mathrm{O^f}$) designates O that is close to (far from) N. Similarly, for the --COOH group, $\mathrm{O^H}$ ($\mathrm{O^n}$) denotes O that does (does not) form a covalent bond with H. And $\mathrm{H^N}$ ($\mathrm{H^O}$) represents the H bonded to N (O).}
  \label{si_fig_6geo}
\end{figure}

\begin{figure}
  \centering
  \includegraphics[width=\textwidth]{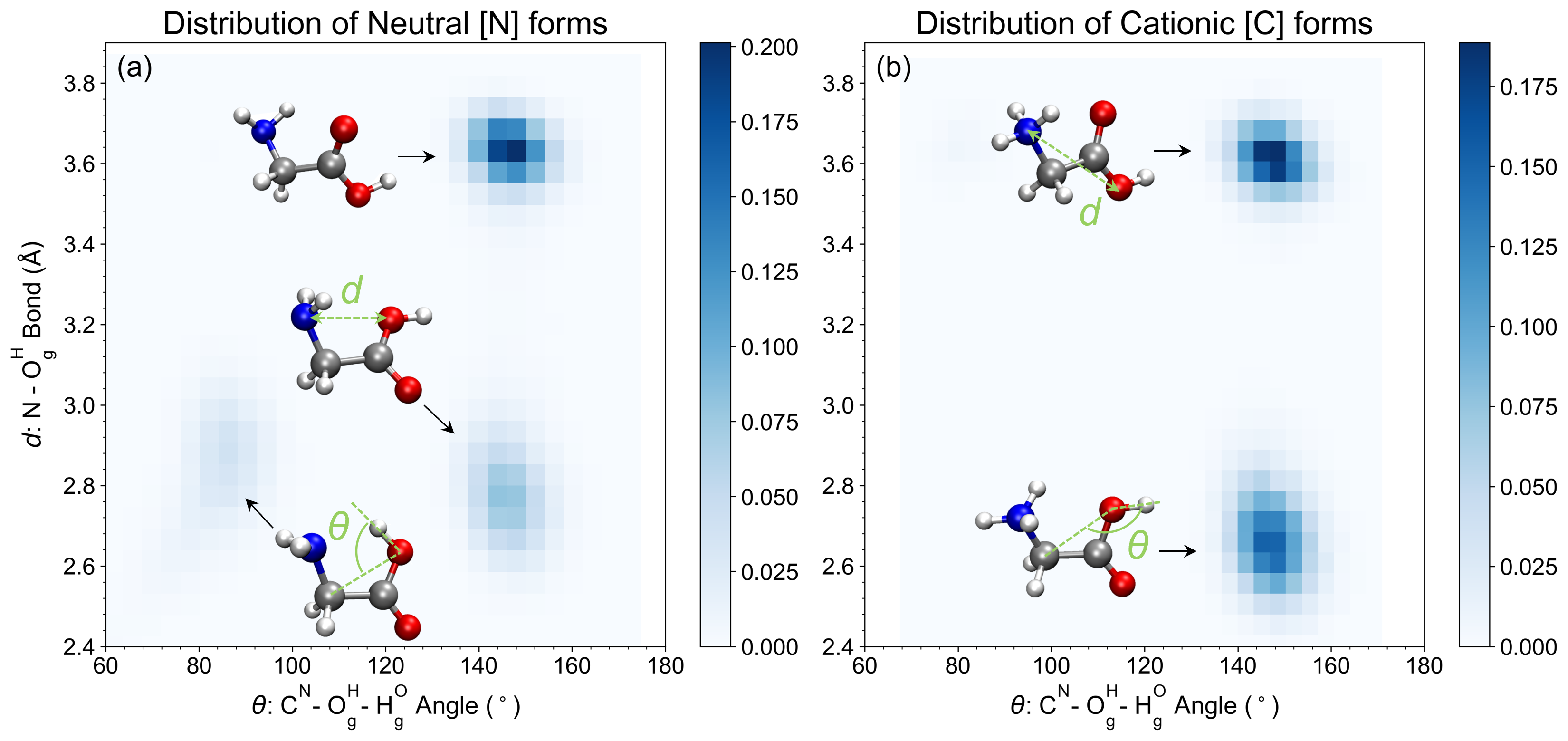}
  \caption{\textbf{Configurational distribution.} The distribution of glycine in (a) neutral [N] and (b) cationic [C] forms as a function of the bond (N-$\mathrm{O^H_g}$) and the angle ($\mathrm{C^N}$-$\mathrm{O_g^H}$-$\mathrm{H_g^O}$) as shown in the schematic diagram of molecular structures. The color bars depict the density of configuration numbers.}
  \label{si_fig_bond_angle_N_C}
\end{figure}

\begin{figure}
  \centering
  \includegraphics[width=\textwidth]{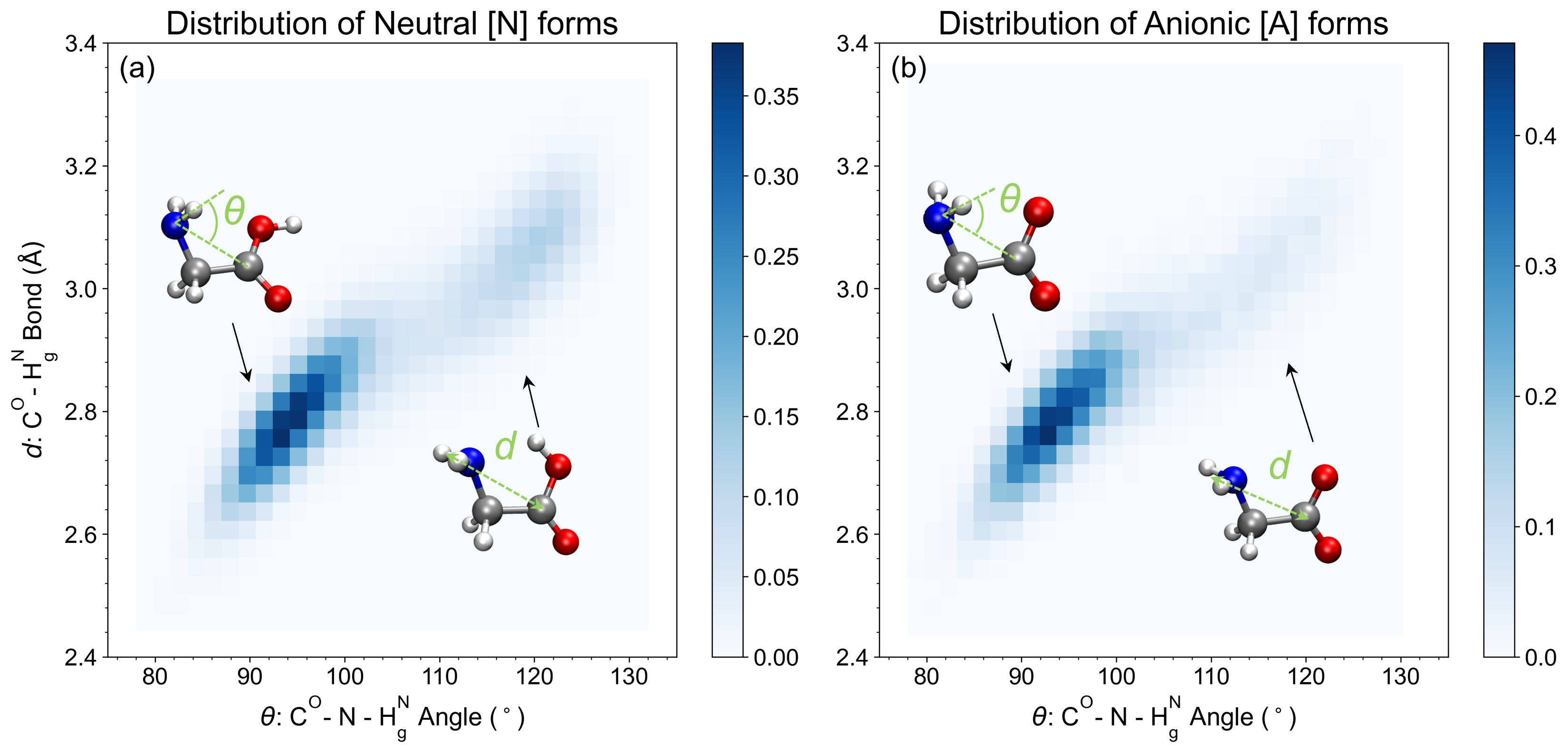}
  \caption{\textbf{Configurational distribution.} The distribution of glycine in (a) neutral [N] and (b) anionic [A] forms as a function of the bond (the average of two $\mathrm{C^O}$-$\mathrm{H_g^N}$ bond lengths) and the angle (the average of two $\mathrm{C^O}$-N-$\mathrm{H_g^N}$ angles) as shown in the schematic diagram of molecular structures. The color bars depict the density of configuration numbers.}
  \label{si_fig_bond_angle_N_A}
\end{figure}

\begin{figure}
  \centering
  \includegraphics[width=\textwidth]{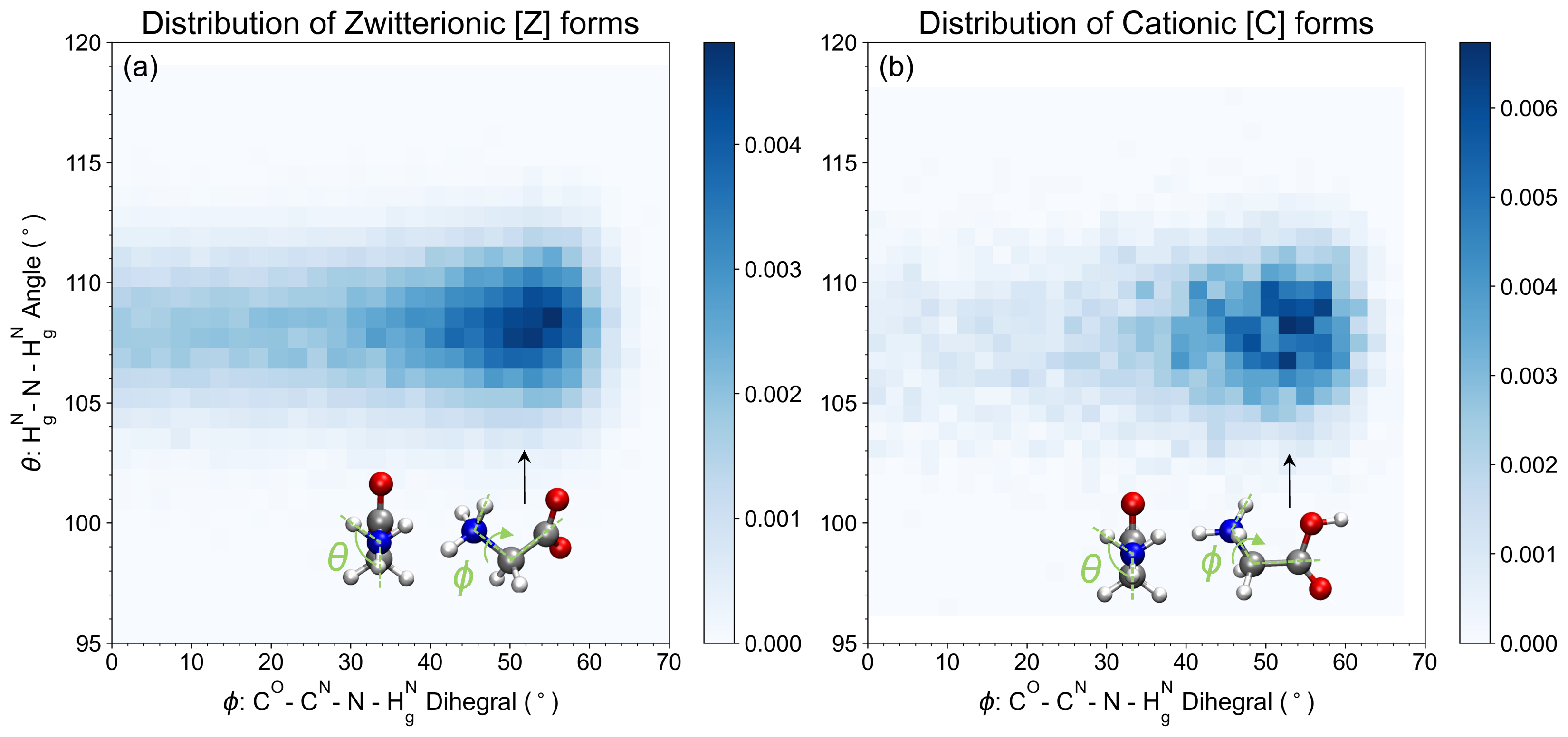}
  \caption{\textbf{Configurational distribution.} The distribution of glycine in (a) zwitterionic [Z] and (b) cationic [C] forms as a function of the angle (the average of three $\mathrm{H_g^N}$-N-$\mathrm{H_g^N}$ angles) and the dihedral (the minimum of three $\mathrm{C^O}$-$\mathrm{C^N}$-N-$\mathrm{H_g^N}$ dihedrals) as shown in the schematic diagram of molecular structures. The color bars depict the density of configuration numbers.}
  \label{si_fig_angle_digedral_Z_C}
\end{figure}

\begin{figure}
  \centering
  \includegraphics[width=0.65\textwidth]{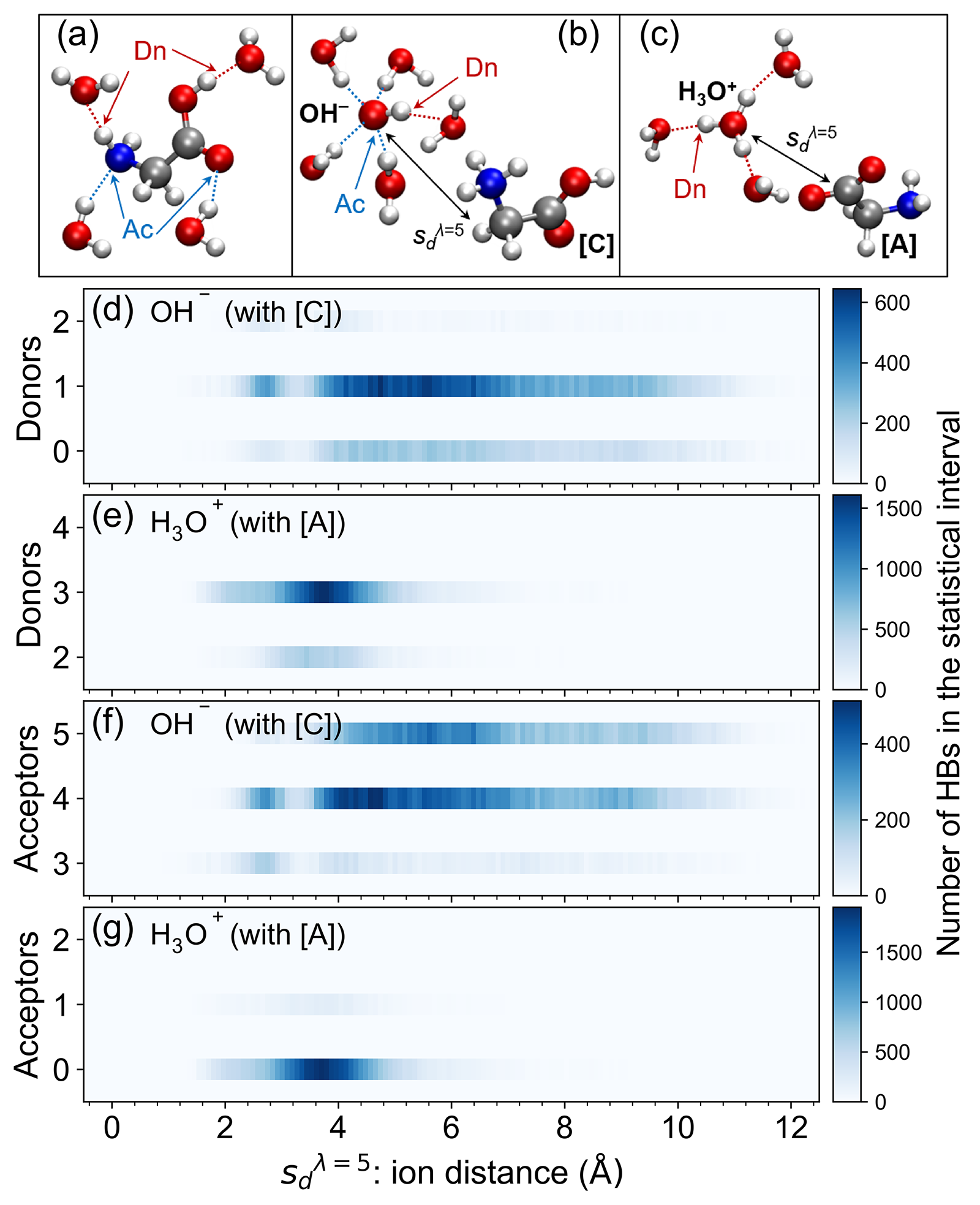}
  \caption{\textbf{Hydrogen bond of self-ions}. The schematic representation of (a) donor (Dn) and acceptor (Ac) sites in the neutral [N] form. The presence of (b) hydroxide ($\mathrm{OH^-}$) with [C] form, and (c) hydronium ($\mathrm{H_3O^+}$) with anionic [A] form in water. The number of donor for (d) $\mathrm{OH^-}$ and (e) $\mathrm{H_3O^+}$, and the number of acceptor for (f) $\mathrm{OH^-}$ and (g) $\mathrm{H_3O^+}$ are displayed as a function of charge-charge distance ($\mathbf{s}_d^{\lambda=5}$).}
  \label{si_fig_Ion_HB}
\end{figure}

\clearpage
\renewcommand{\bibfont}{\footnotesize\linespread{0.9}\selectfont}
\bibliography{references}